\documentclass[12pt,preprint]{aastex}

\usepackage{graphicx}
\usepackage{subfigure}

\newcommand{\etal}{\mbox{et~al.}}
\newcommand{\eg}{e.g.\@~}

\newcommand{\fig}{Fig.~}
\newcommand{\tbl}{Tbl.~}

\newcommand{\kms}{km\,s$^{-1}$} 

\shorttitle{FGLR distance to M\,33}
\shortauthors{U, Urbaneja, Kudritzki \etal}

\begin{document}

  \title{A New Distance to M33 Using Blue Supergiants and the FGLR Method}
  
  \author{Vivian U, Miguel A. Urbaneja, Rolf-Peter Kudritzki, Bradley A. Jacobs, Fabio Bresolin}
  
  \affil{Institute for Astronomy, University of Hawaii,
    2680 Woodlawn Drive, Honolulu, HI 96822}

  \email{vivian@ifa.hawaii.edu,urbaneja@ifa.hawaii.edu,kud@ifa.hawaii.edu,
    bjacobs@ifa.hawaii.edu,bresolin@ifa.hawaii.edu}

  \and

  \author{Norbert Przybilla}

  \affil{Dr. Remeis-Sternwarte Bamberg, Sternwartstrasse 7, D-96049
    Bamberg, Germany}
    
  \email{przybilla@sternwarte.uni-erlangen.de}

  \begin{abstract}
The quantitative spectral analysis of medium resolution optical spectra of A and B 
supergiants obtained with DEIMOS and ESI at the Keck Telescopes
is used to determine a distance modulus of 24.93$\,\pm\,$0.11~mag for the 
Triangulum Galaxy M33. The analysis yields stellar effective temperatures, 
gravities, interstellar reddening, and extinction, the combination of
which provides a distance estimate via the Flux-weighted
Gravity--Luminosity Relationship (FGLR). This result is based on an
FGLR calibration that is continually being polished.  An average reddening of
$<E(B-V)>\,\sim\,$0.08 mag is found, with a large variation ranging from
0.01 to 0.16 mag however, demonstrating the importance of 
accurate individual reddening measurements for stellar distance
indicators in galaxies with evident signatures of interstellar
absorption. The large distance  
modulus found is in good agreement with recent work on eclipsing binaries, 
planetary nebulae, long period variables, RR Lyrae stars, and also
with HST observations of Cepheids, 
if reasonable reddening assumptions are made for the Cepheids. Since
distances based on the tip of the red giant branch (TRGB) method found in 
the literature give
conflicting results, we have used HST ACS $V$- and $I$-band images of
outer regions of M\,33 to determine a TRGB distance of
24.84$\,\pm\,$0.10~mag, in basic agreement with the FGLR result. 
We have also determined stellar metallicities and discussed the metallicity 
gradient in the disk of M33. We find metallicity of $Z_\odot$ at the
center and 0.3\,$Z_\odot$ in the outskirts at a distance of one
isophotal radius. The average logarithmic metallicity gradient is
-0.07$\,\pm\,$0.01\,dex kpc$^{-1}$. However, there is a large scatter around this
average value, very similar to what has been found for  
the \ion{H}{2} regions in M\,33.
 \end{abstract}


  \section{Introduction}
  \label{Introduction}

  Modern astronomy has advanced to a stage where cosmological parameters
  are claimed to be determined with hitherto unknown
  precision~\cite[\eg][]{Frieman08}. At the same time, however, we are 
  confronted with the fact that distances to many nearby galaxies are 
  disturbingly uncertain. An interesting and important example is the
  distance to the Triangulum Galaxy M\,33 (see~\citealt{Bonanos06} for a
  summary). Several independent distance determination techniques have
  been employed for this galaxy since the beginning of this decade,
  including Cepheids and RR
  Lyrae~\cite[][]{Freedman91,Lee02,Sarajedini06}, tip of the red 
  giant branch (TRGB) and red clump
  stars~\cite[][]{Rizzi07,Galleti04,McConnachie04,Tiede04,Kim02}, 
  planetary nebulae~\cite[][]{Ciardullo04}, horizontal branch
  stars~\cite[][]{Sarajedini00}, long period variables~\cite[][]{Pierce00},
  detached eclipsing binary~\cite[][]{Bonanos06}, 
  and water masers~\cite[][]{Brunthaler05}. The shocking result is that the 
  distance moduli obtained with these different methods differ by as much 
  as 0.6 magnitudes, which is more than 30 percent in linear distance. Even 
  when the water maser distance were discounted in this comparison as
  justified by its large intrinsic uncertainty of 0.45 mag, the spread
  in distance moduli is still about 0.4 magnitudes, significantly higher than 
  the uncertainties quoted in the individual works.

  A significant part of the uncertainty comes from interstellar extinction. 
  M\,33 is a mild-inclination star-forming galaxy with a patchy dusty disk that exhibits
  inhomogeneous intrinsic extinction. This has a considerable effect for many of 
  the aforementioned methods. In particular, distance
  determination using Cepheids at $V$ and $I$ bands is severely affected,
  as they are part of a young stellar population in a dusty star-forming 
  environment. As pointed out by~\cite{Kim02} and~\cite{Ciardullo04}, the difference between 
  the shorter distances obtained with Cepheids and the longer distances found with 
  other methods would disappear given smaller reddening than assumed in the 
  Cepheid work cited above. Interestingly, the work by~\cite{Bonanos06}, 
  which uses quantitative spectroscopy of a detached eclipsing binary and, thus, 
  comes up with an accurate determination of reddening and extinction, yields a large
  distance modulus of 24.92\,mag, 0.3 magnitudes larger than the one obtained with $V$- and 
  $I$-band photometry of Cepheids.

  The solution to resolving this discrepancy is to use a method that
  is either unaffected by reddening and extinction uncertainties such
  as K-band photometry of Cepheids~\cite[\eg][]{Gieren05}, or one that
  includes the accurate determination of reddening for each individual
  star gauged in the determination of the distance modulus. An example
  for the latter involves the quantitative spectroscopy of massive
  blue supergiant stars and the use of the Flux-weighted
  Gravity--Luminosity Relationship \cite[FGLR;][]{Kudritzki03} . In
  this paper we will apply this method to determine the reddening,
  extinction, and distance to M\,33. 
  
  Blue supergiants of B and A spectral types are the optically brightest 
  stars in the universe with visual magnitudes up to $M_{V} \cong -9.5$\,mag. Their 
  quantitative spectral analysis based on state-of-the-art non-LTE
  radiative transfer models allows for precise determination of
  stellar parameters, metallicity, reddening, and 
  extinction~\cite[][]{Urbaneja05b,Przybilla06,Kudritzki08}. Then,
  with stellar temperatures $T_{\rm eff}$ and gravities $\log\,g$, one
  can use the de-reddened apparent stellar magnitudes and the tight
  relationship between flux-weighted gravity 
  $\log\,g_{F}\,=\,\log\,g\,-\,4\,\log\left(T_{\rm eff} \times
    10^{-4}\right)$ and absolute bolometric magnitude  
  $M_{\rm bol}$, to determine a distance via the FGLR. 

  The concept of FGLR has been introduced by~\citet{Kudritzki03}. It is based 
  on the assumption that massive stars evolve through the
  Hertzsprung-Russell (H-R) diagram temperature domain of 
  B and A spectral types at constant luminosity and mass. In a detailed spectroscopic 
  study of blue supergiants in the nearby spiral galaxy NGC\,300, which is at a distance
  of 1.9\,Mpc, \citet[][hereafter K08]{Kudritzki08} verified this concept and
  contributed towards the calibration of the FGLR. \cite{Urbaneja08} 
  then applied the FGLR method and determined the distance to the
  metal-poor, irregular Local Group galaxy WLM, the result from which
  agreed well with most recent distance determinations based on K-band
  photometry of Cepheids and I-band photometry of the TRGB.

  Motivated by the success of previous application of the FGLR as well
  as the prospects of this work, we undertake an independent
  investigation of the distance to M33 by means of a quantitative
  analysis of medium resolution optical spectra of blue supergiant stars. 

  This paper is organized as follows:  In \S \ref{Data} we describe
  the data sample used to obtain the FGLR of M33 as well as the 
  observations undertaken at Keck Observatory.  In \S
  \ref{Spectral Analysis} we detail the process of the quantitative spectroscopic 
  analysis. We present the results of the spectral analysis in \S \ref{Results} and 
  discuss stellar parameters and stellar evolution, metallicity and metallity gradient, 
  and reddening and extinction. This provides the basis for a new distance determination 
  using the FGLR in \S \ref{Distance}. A discussion of the new result compared with 
  distances determined from other methods is given in \S \ref{Discussion}. Finally, we 
   present our conclusions and future outlook in \S \ref{Conclusions}.  

 \section{Observations}
 \label{Data}

Distance determination using the FGLR method requires a sufficiently large 
sample (10 to 20 objects, see K08, \citealt{Urbaneja08}) of supergiants with 
good multi-band photometry and spectra of reasonable quality that are suitable
for quantitative analysis using model atmosphere and radiative
transfer techniques. We pre-selected a sample of such potential
targets that fulfill the initial color-magnitude criteria described in
K08.  \cite{Massey06} provides an excellent source for photometry of
all these objects; spectra were collected with the Keck Telescopes on
Mauna Kea. In our first observing run for this project on November 1,
2003 with good seeing ($0\farcs6$), we used the DEIMOS spectrograph
with a 1200 l/mm grating yielding a dispersion of 0.33~\AA/px and a
spectral resolution (FWHM) of 1.6~\AA. The total integration time was
135 and 93 minutes for the blue and red spectra, respectively. A total 
number of 49 targets were observed with one DEIMOS mask.~Among 
these, we selected 10 objects with
spectral type ranging from B3 to A3 based on the requirement of a
minimum signal-to-noise ratio (SNR) needed to carry out the ensuing spectral analysis. 

Additional observations of 7 more A-type supergiants were
carried out under good weather conditions (clear sky, typical seeing
around $1\farcs$0) on October 8, 2005 and October 4, 2007 with the ESI
spectrograph \cite[][]{Sheinis02}. The instrument was used in
echellette mode, with a slit width of $0\farcs75$, providing
R$\,\sim\,$9000 (equivalent to 0.6 \AA~FWHM at 5500 \AA).   
Depending on the visual magnitude of the object, one to three
exposures were acquired, with the exposure time adjusted to provide a
final SNR$\,\sim\,$100 in the continuum.
Unlike in the case of DEIMOS, the ESI stars were selected to 
cover a wide range of galactocentric distances.
  
To further increase the sample size for FGLR analysis and calibration, we included
additional stars from the spectroscopic study by~\cite{Urbaneja05}. In this work,
11 B supergiants of early spectral type (B0 to B2) were analyzed quantitatively using 
state-of-the-art non-LTE model atmospheres to investigate the M33 metallicity gradient. 
While these stars were well suited for the abundance study, however, care must be taken 
while considering their selection as targets for the FGLR distance determination method. Two of 
them (B133 and 1054) were found to be multiple systems with companions
that would affect magnitudes and colors. Two others (110A and OB10-10) turned out
to be extreme blue hypergiants, with very high luminosities, rendering the determination
of their surface gravities less reliable. For object 1137, only an upper limit for the
mass loss rate could be found, which means that, because of wind
contamination in the Balmer lines, only an upper limit for the gravity
could be determined. The spectra for object 0900 suffered from strong
\ion{H}{2} region emission and consequently-distorted profiles of the
two observed Balmer lines
H$\gamma$ and H$\delta$, again leading to a rather uncertain
stellar gravity. The remaining five objects have very well determined 
stellar parameters and have been incorporated as targets in the FGLR
distance determination technique.  This brings the total number of M\,33
supergiants in our sample up to 22. The list of all the
objects is given in~\tbl \ref{tbl:id}.

\begin{figure}[pth]
    \centering
    \subfigure{\label{fig:idl21259}
      \includegraphics[angle=90,height=0.4\textheight]{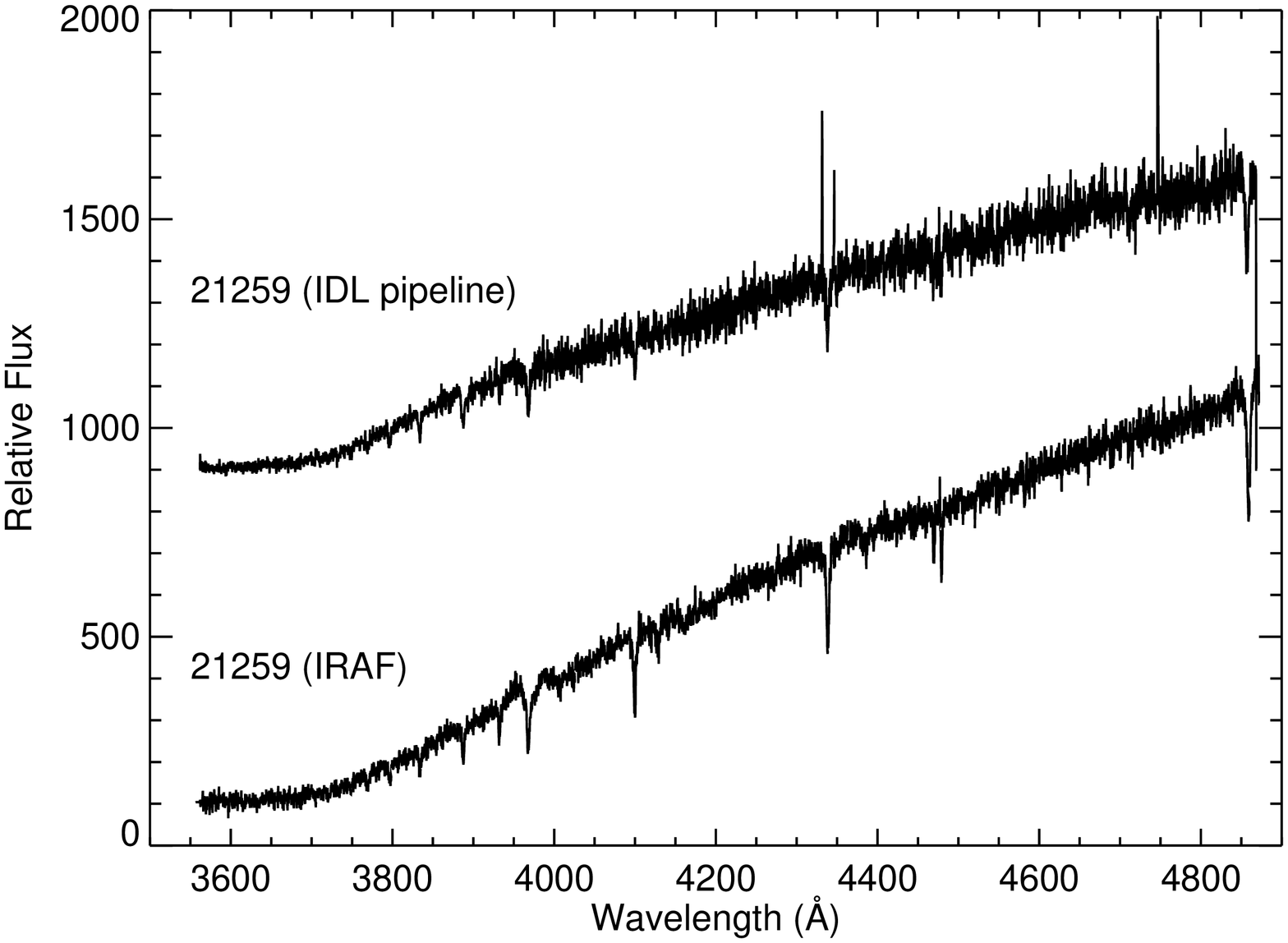}}
    \hspace{.1in}
    \subfigure{\label{fig:idl767}
      \includegraphics[angle=90,height=0.4\textheight]{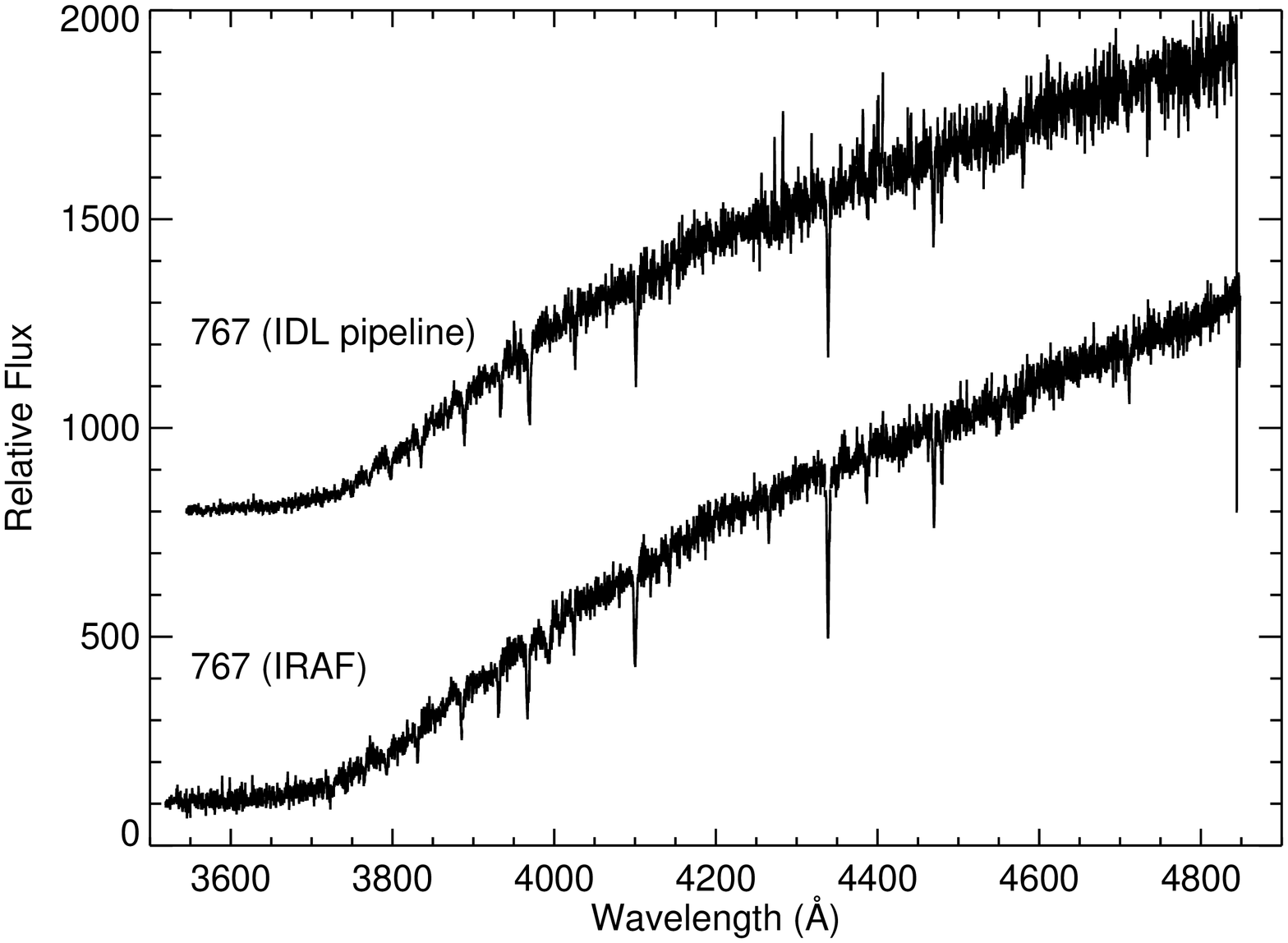}}
    \caption{Comparison of pre-normalized flux spectra for two selected
	supergiants of late (top) and early (bottom)
	 B types reduced by DEEP2 IDL pipeline
	and by IRAF reduction packages, respectively.  Notice that the 
	SNR in this blue wavelength range is much better for the IRAF-reduced
	spectra than for the pipeline results.}
    \label{fig:comp_idl}
  \end{figure}

The new ESI data were reduced in the same way as described in~\cite{Urbaneja05}.
On the other hand, the DEIMOS spectra were initially reduced using the DEEP2 Reduction 
Software~\cite[\eg][]{Marioni01}, but the resulting SNR in the short
wavelength range were insufficiently low because the subroutines in
the pipeline optimizes in the longer wavelength regime. Therefore, the data reduction
of the DEIMOS data was carefully redone step-by-step using the
standard Image Reduction and Analysis Facility (IRAF) packages. As
demonstrated in \fig\ref{fig:comp_idl}, the final extracted spectra
have a clearly improved SNR in the blue region, giving better results
than the DEEP2 pipeline did. The noticeably-improved SNR allows us to 
identify and measure the relevant spectral features more precisely.  However, even
after this improvement in SNR, the poor blue sensitivity of DEIMOS (as can be 
seen in  \fig\ref{fig:comp_idl}) turned out to be an extra challenge during the 
spectral analysis. 

 \section{Spectral Analysis}
 \label{Spectral Analysis}

The analysis method to be applied to the individual targets depends on
the targets' spectral types. For early B supergiants, we use the 
Si~{\sc ii/iii/iv} equilibrium based on metal-line-blanketed non-LTE model
atmospheres and line formation calculations that include
effects of stellar winds and spherical extension to constrain effective 
temperatures~\cite[see][]{Urbaneja05b,Urbaneja05}. For late B
and early A spectral types, the determination of effective temperature
utilizes the Balmer jump or ionization equilibrium such as Mg~{\sc i/ii} 
and O~{\sc i/ii}.  These models involve LTE
line-blanketed model atmospheres in hydrostatic equilibrium with
plane-parallel geometry and very detailed non-LTE line formation calculations 
\cite[][]{Przybilla06,Kudritzki08}. In consequence, as a first step
towards quantitative spectral analysis, we determined spectral types
for all 14 newly-observed objects through a qualitative comparison
with a complete set of Milky Way and SMC supergiant template spectra (see
\citealt{Bresolin01,Lennon97}). Only three objects were found to exhibit
early spectral type of B3 while that of the rest ranges between B8 to A2 
(see \tbl \ref{tbl:summary}).   

Spectra analyzed in the FGLR studies by K08 and \cite{Urbaneja08} had a
rather low resolution of 5$\,$\AA. While this resolution is
sufficiently good for determining stellar gravities from the Balmer
lines, it precludes the use of ionization equilibria for temperature
determination in the case of late B and early A supergiants because
the weak spectral lines of the neutral species disappear in the noise
of the spectra. As an alternative, the Balmer jump or the strengths of
the neutral helium lines were utilized instead. However, with the
spectroscopic resolution in our Keck DEIMOS and ESI
spectra (1.6 and 0.6 \AA~respectively), we can employ the 
technique of ionization  equilibria in this work~\cite[see][]{Przybilla06}. 

The basis for the spectral analysis of late B and A type supergiant
stars is the grid of model atmospheres and line formation  
calculations in K08 that spans an extensive area in 
($T_{\rm eff}$,~$\log~g$) parameter space (see \fig 1 in K08) and
covers a large range in metallicity from $0.05\,Z_\odot$ to
$2\,Z_\odot$. We refer to K08 for a description of the computed grid
(abundances, choice of microturbulence velocities, etc.).
The input physics of the model atmospheres and line formation 
calculations are detailed in~\cite{Przybilla06}. Regarding the mid
B-types, we use the unified non-LTE model atmosphere/line formation
code {\sc fastwind} \cite[][]{Puls05} to create a small grid of
models, with the parameters properly selected to explore the expected
small parameter space covered by these three objects. 

The analysis proceeds in three steps. First, projected rotational velocity is
determined; second, stellar effective temperature and gravity; third, stellar 
metallicity. If needed, steps two and three are iterated. In the following, we 
describe the individual steps.

 \subsection{Rotational Velocities} 
Projected rotational velocities $v\sin\,i$ of late B and early A supergiants 
are usually between 30 to 70 \kms~\cite[][]{Howarth04}
comparable to or
somewhat larger than the resolution of the DEIMOS and ESI  
spectrographs, respectively. In other words, we have to determine the amount of 
rotational line broadening in addition to instrumental broadening for each individual 
star before we start the spectral analysis via comparing synthetic with observed 
line profiles.

\begin{figure}
    \centering
    \includegraphics[angle=90,width=1.1\textwidth]{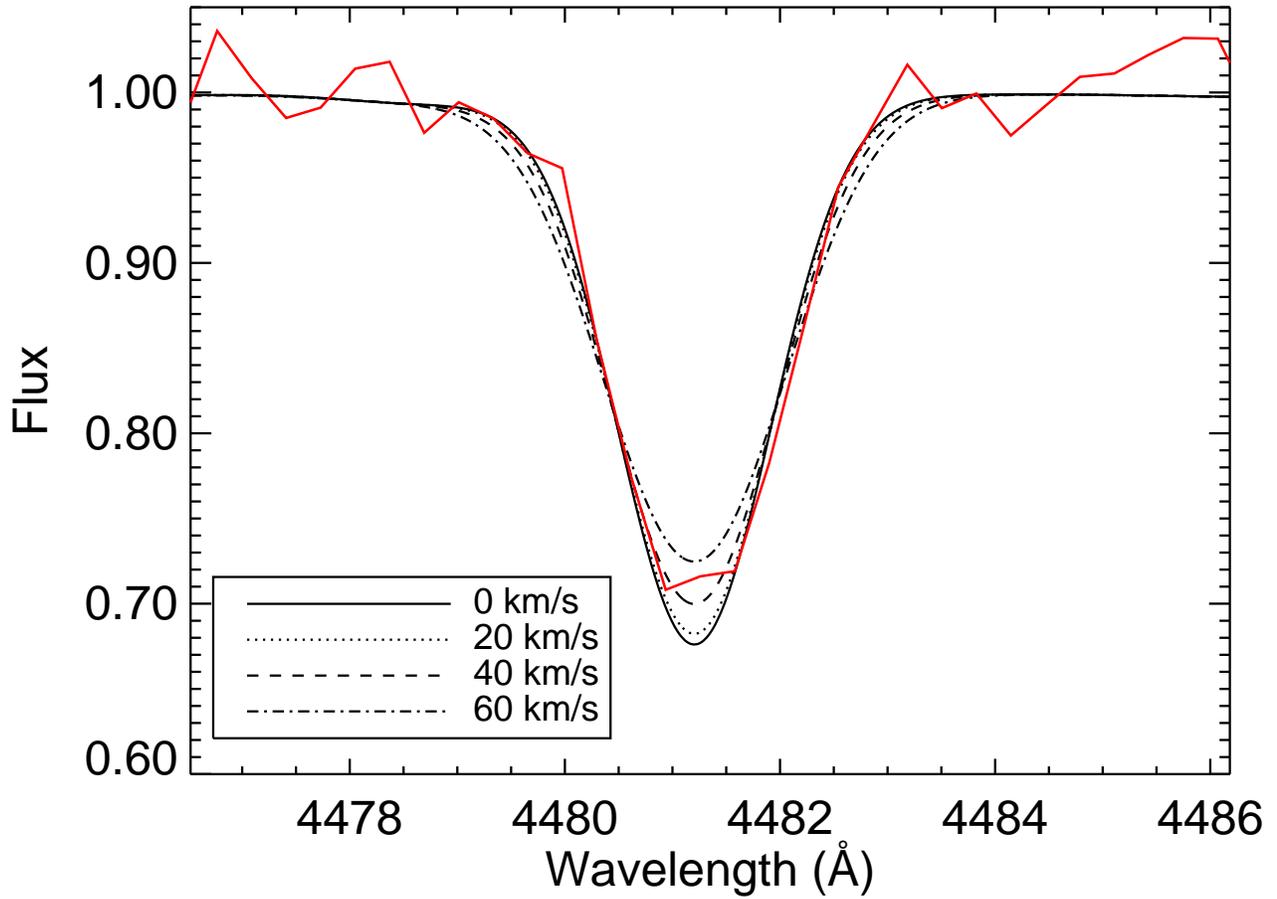}
    \caption{The effects of different rotational velocities on the shape of the
	\ion{Mg}{2} (4481) line.  As the velocity increases, the line grows flatter and
	flatter, with more pronounced effects in the core than in the wings.}
    \label{fig:comp_vrot}
  \end{figure}

With the large number of objects to be analyzed, it is important to
have a simple method to accurately estimate $v\sin\,i$ that is
independent of stellar parameters as well as the details of the final
atmospheric model synthesizing the observed spectrum. A good approach
is to look at lines that are weak enough to not suffer from pressure
broadening to a large degree but still sufficiently strong such that
the rotation-broadened profile shape can be well
determined~\cite[][]{Gray92}. The \ion{Mg}{2} $\lambda$4481 line, the
\ion{Si}{2} $\lambda$6347, 6371 lines and the
\ion{Fe}{2} $\lambda$4515, 4535 lines are ideal candidates for late B
through early A supergiants while the \ion{Si}{3} lines at
$\lambda$4552, 4567, 4574 are useful for early B supergiants. For
those lines we construct a Gaussian profile with very narrow FWHM
(0.1~\AA) that has the same equivalent width as the observed line. We then convolve 
this initial profile with the instrumental profile and a rotational profile, where 
we steadily increase $v\sin\,i$. An example in the case of the DEIMOS spectra is given in 
\fig \ref{fig:comp_vrot}. A $\chi^{2}$ minimization of the difference between the 
observed and calculated profiles in turn indicates projected
rotational velocities.  We note a caveat that \ion{Mg}{2}
$\lambda$4481 is a blend of two strong components separated by 0.2\,\AA;
determination of $v\sin\,i$ with a single Gaussian may tend to
overestimate $v\sin\,i$ in these cases.

The lower spectral resolution and SNR achieved for the  DEIMOS data
do not allow us to disentangle rotation and macroturbulence. Therefore, the value of $v\sin\,i$ 
thus obtained incorporates macroturbulence and might overestimate the true rotational 
velocity. This, however, will not affect the determination of stellar parameters and 
subsequent FGLR distance analysis. The situation is more optimistic
for the late B and early A supergiants observed with ESI that have
significantly better spectral resolution and SNR; we thus decided to
apply a two-step procedure. First, $v\, \sin i$ was determined by a
Fourier transform of individual lines (such as the ones mentioned  
previously) as described in~\citet{Gray92}. With this value fixed, the macroturbulence 
velocity was then established by a by-eye fit of the observed profile similar 
to that done in~\citet{Przybilla06}. The early B supergiant sample
from \cite{Urbaneja05} were in a situation similar to that of our
DEIMOS spectra: in order to obtain a reasonable SNR, the ESI spectra
had to be binned to the detriment of the resolution, which did not allow then to 
discern between rotation and macroturbulence. 

A list of rotational and macroturbulence velocities for our sample of supergiants is 
given in \tbl \ref{tbl:summary}. The uncertainties are on the order of 5 \kms.

  \subsection{Gravity and Effective Temperature}
The standard technique for determining gravities and effective temperatures of late B 
and early A supergiants is described in detail by~\cite{Przybilla06} and K08. We 
apply a very similar but somewhat more efficient technique in this work. 

\begin{figure}
    \centering
    \includegraphics[]{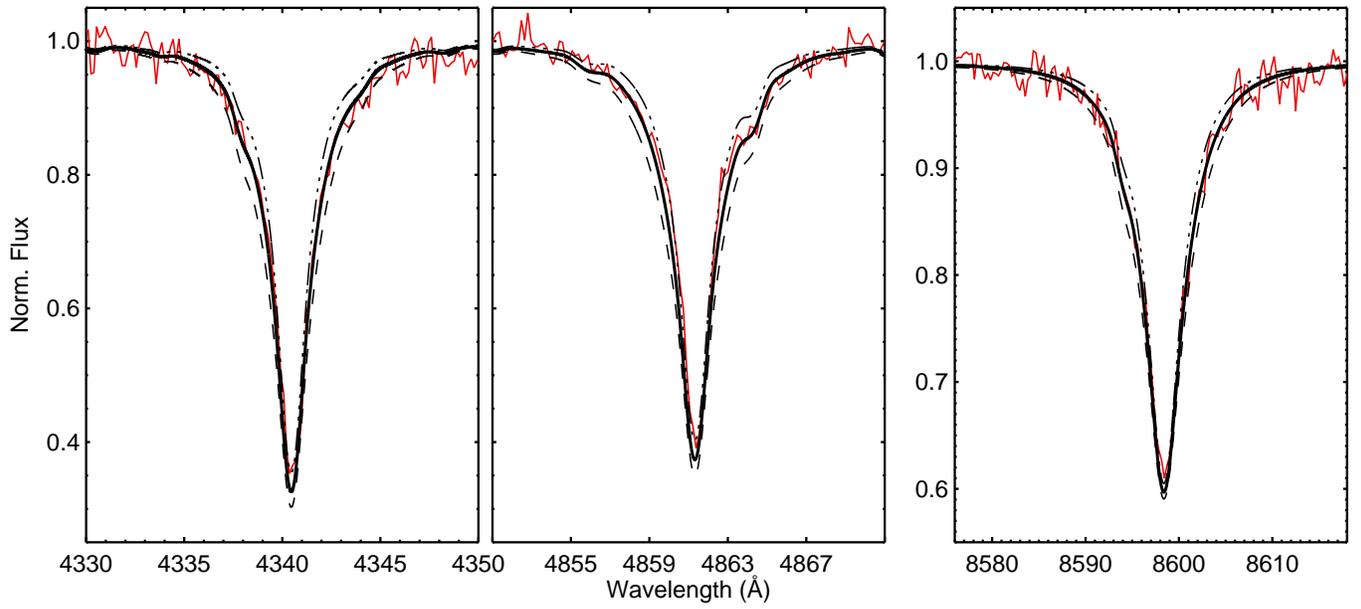}
    \caption{These panels show the Balmer lines fit (from left to right: 
      H$_\gamma$, H$_\beta$, H$_{3-14}$) for the object No.\,11.  The observed spectrum
      (thin) is overplotted with the model (thick) with $T_{eff}$ of 9600\,K and
      $\log\,g$ at 1.40$^{+0.10}_{-0.10}$ dex where the errors are represented by dashed lines.      
      \label{fig:gravfit_new_1} }
  \end{figure}

\begin{figure}
  \centering
  \subfigure{\label{fig:g_fit}
    \includegraphics[angle=90,width=0.8\textwidth]{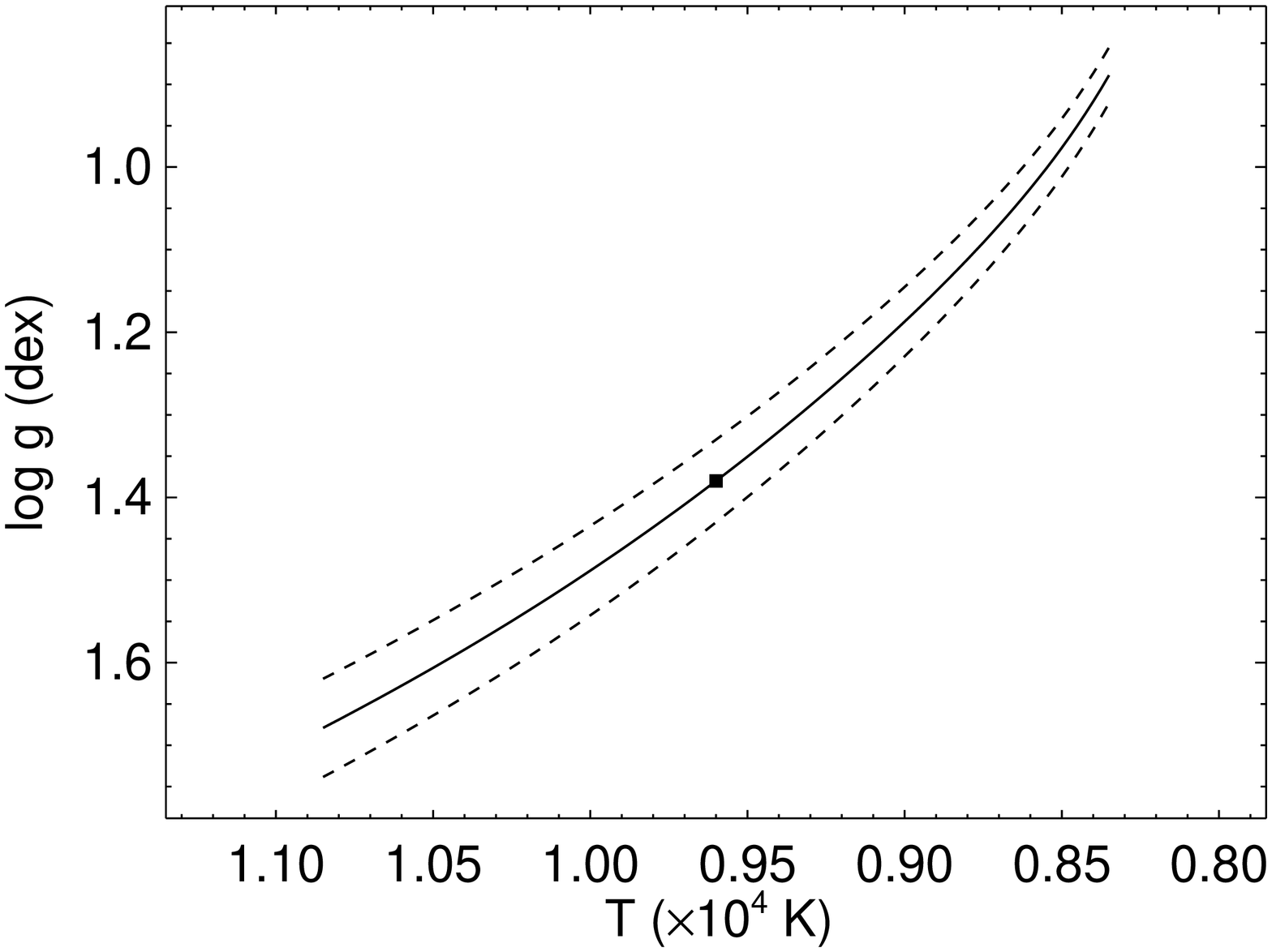}}    
  \hspace{.1in}
  \subfigure{\label{fig:gf_fit}
    \includegraphics[angle=90,width=0.8\textwidth]{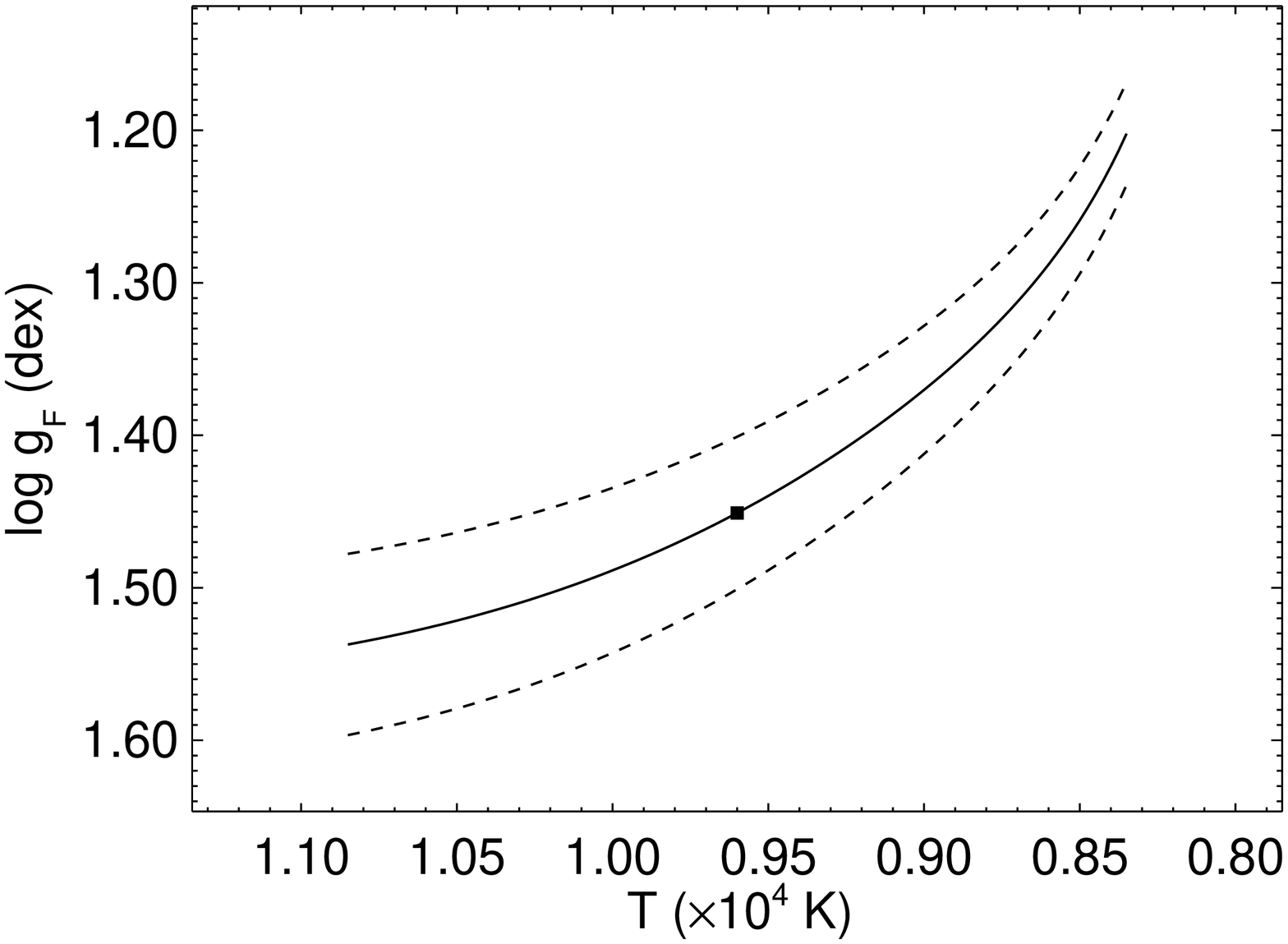}}
  \caption{Balmer line fit curve for object No.\,11 in the
    $\log\,g$--$T_{\rm eff}$ plane 
    (top) and the $\log\,g_{F}$--$T_{\rm eff}$ plane (bottom). 
    The dashed curves correspond to the maximum fitting errors.
  } 
  \label{fig:gravfit_new_2}
\end{figure}

The first step involves using information about gravity and effective temperature 
that are provided by a fit of the Balmer line profiles. Balmer lines 
are an excellent indicator of stellar gravity because of their strong density dependence 
on pressure broadening through the Stark effect. An example is given in 
\fig \ref{fig:gravfit_new_1} for object No.\,11 of our sample. At a fixed effective 
temperature, the step size of each gravity fit on the grid is 0.05 dex. Since the excitation of 
the second level of hydrogen and the ionization of hydrogen is temperature-dependent, 
the strength of the Balmer lines also depends on effective temperature. This 
means that the Balmer lines may be fitted just as well as higher
(lower) effective temperature and higher (lower) gravity than shown for 
the example in \fig \ref{fig:gravfit_new_1}. The corresponding fit curve on the 
($\log~g, T_{\rm eff}$)-plane is demonstrated in \fig \ref{fig:gravfit_new_2}. Since the major 
motivation for this work is a determination of the distance to M\,33 using the stellar 
flux-weighted gravity $\log g_{F}$, we also plot the same fit curve for the Balmer lines 
on the ($\log~g_{F}$, $T_{\rm eff}$)-plane. We note that 
for stars sufficiently hot, $\log\,g_{F}$ can be solely determined by fitting the Balmer 
lines. The physics behind this phenomenon is discussed in detail in K08.

\begin{figure}
    \centering
    \includegraphics[angle=90,width=0.9\textwidth]{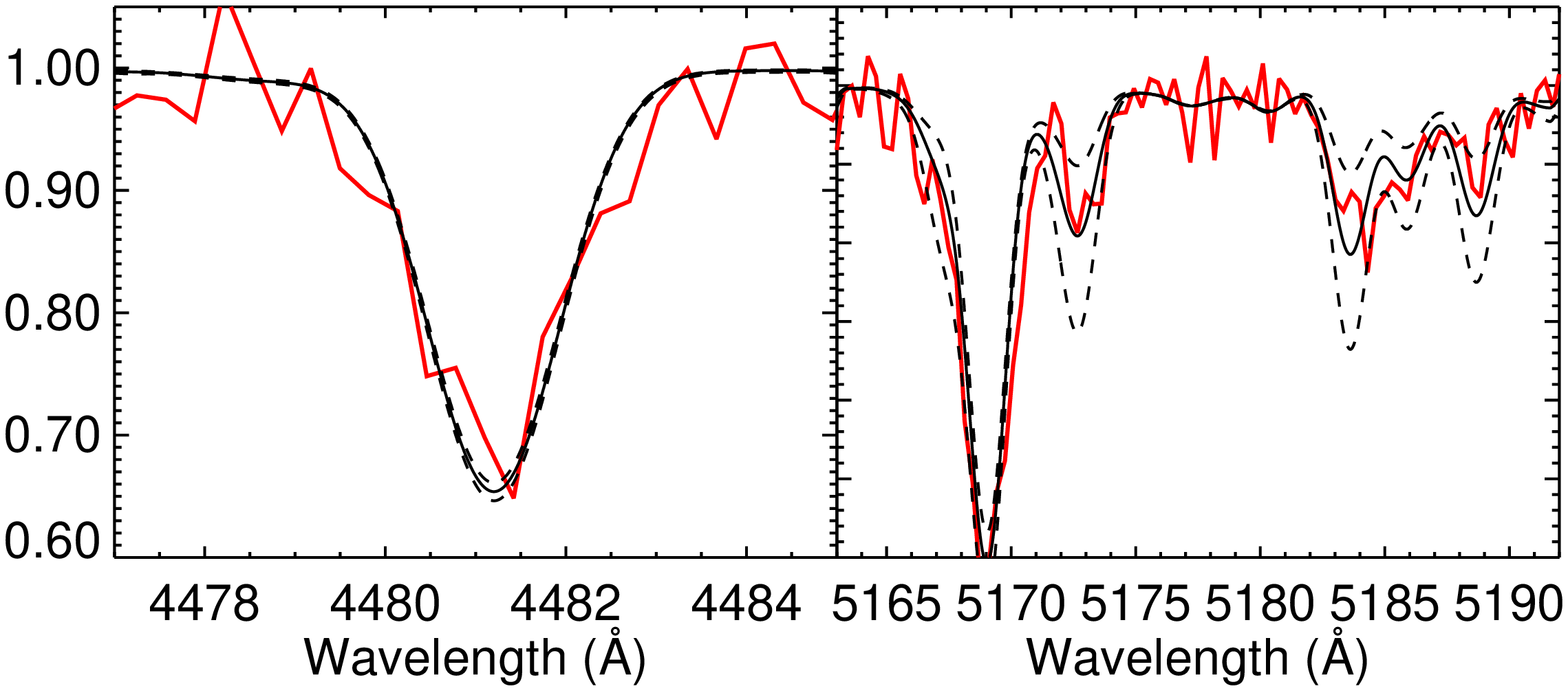}
    \caption{Fit of the Mg I/II ionization equilibrium along the Balmer line
      fit curve of object No. 5. Three models are overplotted.
      $T_{\rm eff}$ = 8750K and $\log~g$ = 1.30 (solid curve), $T_{\rm
      eff}$ = 8300K and $\log~g$ = 1.05 and $T_{\rm eff}$ = 9250K and
      $\log~g$ = 1.50 (both dashed). While the Mg II line at
      4481\AA~is insensitive to model parameter changes along the fit
      curve in this temperature range and depends only on metallicity,
      the MgI lines (vertical bars at the right hand figure) show a
      very strong temperature dependence. A temperature of $T_{\rm
      eff}$ = 8750$\pm250$ is obtained from this plot.       
}
    \label{fig:tfit_new_1}
  \end{figure}

While the analysis of the Balmer lines provides a strong constraint for gravity
and temperature, we need a second, more temperature-dependent 
spectral feature to further constrain the effective temperature along the Balmer
line fit curve in \fig \ref{fig:gravfit_new_2}. The most accurate solution is to
use ionization equilibria such as Mg~{\sc i/ii} for late B and early A supergiants 
\cite[see][]{Przybilla06} and Si~{\sc ii/iii/iv} for early B supergiants 
\cite[see][]{Urbaneja05}. 
\fig \ref{fig:tfit_new_1} demonstrates how this ionization equilibrium changes 
with ($T_{\rm eff}$, $\log~g$) pairs and sets the best-fit effective
temperature of object No.\,5 to be 8750$\,\pm\,$250 K.

\begin{figure}
    \centering
    \subfigure{\label{fig:he}
      \includegraphics[angle=90,width=0.7\textwidth]{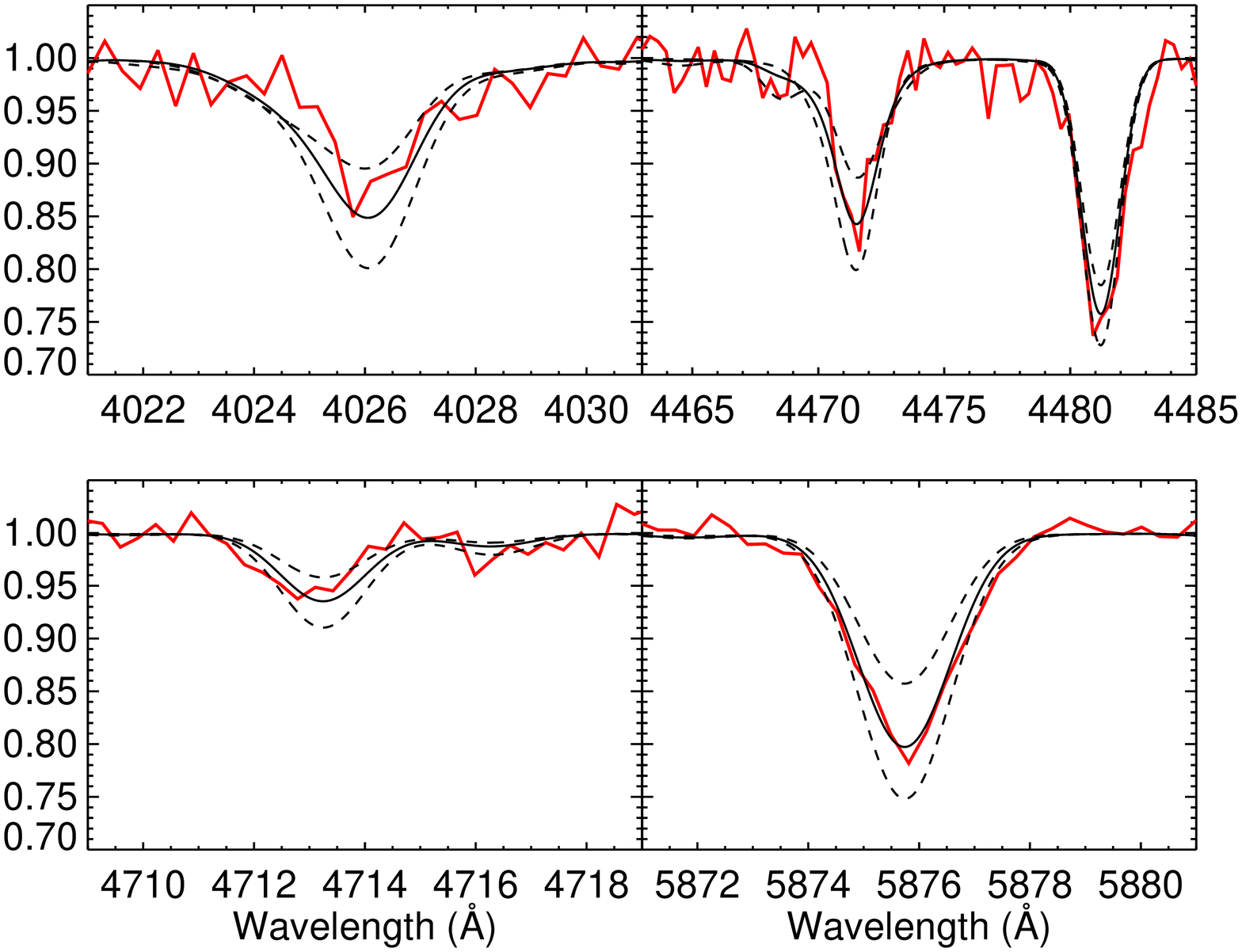}}
    \subfigure{
      \vspace{-0.5in}
      \includegraphics[angle=90,width=0.8\textwidth]{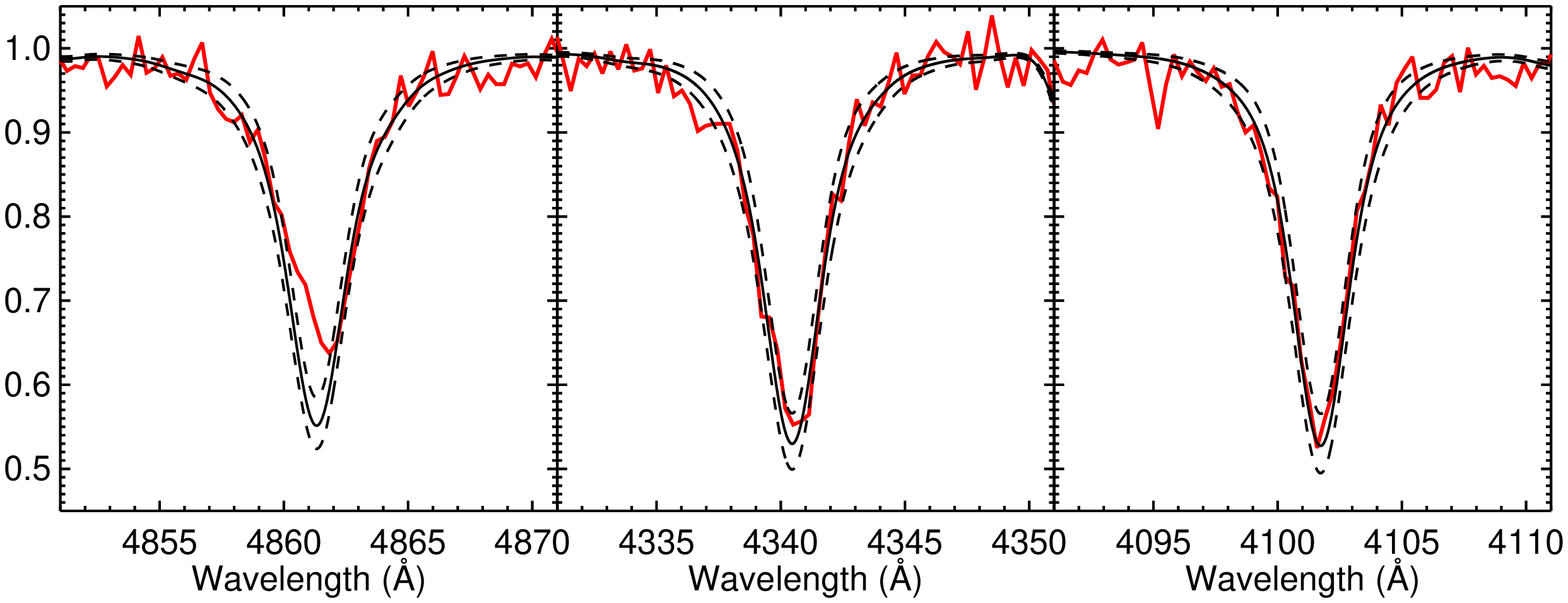}}
    \caption{Line fit of object No. 1. Top: the HeI lines at 4026\AA, 4471\AA, 4713\AA, and 5876\AA
      ~along the Balmer line fit curve of object No. 1. Three models are overplotted.
      $T_{\rm eff}$ = 11000K and $\log~g$ = 1.55 (solid curve), $T_{\rm eff}$ = 10000K and $\log~g$ = 1.35 
      and $T_{\rm eff}$ = 12000K and $\log~g$ = 1.70 (both dashed). Bottom:  Balmer lines (from left to right:
      H$_\beta$, H$_\gamma$, H$_\delta$) at $T_{\rm eff}$ = 11000K. The gravities are $log~g$ = 1.55 (solid)
      and 1.45 and 1.65 (both dashed), respectively. Note that
      H$_\beta$ is affected by stellar wind emission in the line core.   
}
    \label{fig:tfit_new_3}
  \end{figure}

Because of the relatively high resolution and high SNR ($\sim$100), 
the ionization equilibrium method works extremely well for all our targets with 
ESI spectra. The DEIMOS spectra exhibit lower resolution and lower SNR ($\sim\,$30-70),
but the method is still successful for many of the remaining targets,
as shown in \fig \ref{fig:tfit_new_1}. However, there
were some DEIMOS targets for which no \ion{Mg}{1} lines could be observed because 
either the stars were too hot or the spectra were too noisy, or both. For such cases we 
applied an alternative technique introduced by K08 that made use of the 
very strong temperature dependence of the neutral helium lines in the temperature 
range from 9000K to 13000K. An example for such a DEIMOS case is given in \fig
\ref{fig:tfit_new_3}, where a fit of the \ion{He}{1} lines of object
No.\,1 yields $T_{\rm eff}$ = 10000$\,\pm\,$500K. The fit of the 
Balmer lines for the same object is also presented in the figure. 
Naturally, using the \ion{He}{1} lines requires making an assumption
about the helium abundance, which is slightly enhanced in the atmospheres of late B 
and early A supergiants but varies only between $y\,=\,N(He)/(N(He)+N(H))$ = 0.11 to 0.13 
according to the high resolution, high SNR quantitative spectroscopy 
by~\cite{Przybilla06} and \cite{Schiller08}. 
We have adopted y = 0.12 for the fit in \fig \ref{fig:tfit_new_3}. We note, 
though, that the influence of temperature variation on the HeI lines in this temperature 
range is much stronger than, for instance, changing y from the solar value of 
0.09 to 0.13. We also note that for all cases in this temperature range, where we 
were able to constrain the effective temperature with ionization equilibria (\eg 
all our objects observed with ESI), the agreement of the observed
\ion{He}{1} lines and computed models is excellent, lending support
to the reliability of \ion{He}{1} method.

\tbl \ref{tbl:summary} gives the effective temperatures and gravities
thus obtained along with their associated uncertainties.
Note that in most cases the errors for $\log\,g$ are larger than that
for $\log\,g_{F}$.The reason is that the gravity fit curve has a
stronger dependence on temperature than the flux-weighted gravity fit
curve, as illustrated in the upper and lower panels of \fig
\ref{fig:gravfit_new_2}, respectively (see K08 for a detailed
discussion of this effect).

  \subsection{Metallicity}
\begin{figure}
    \centering
    \includegraphics[width=0.9\textwidth]{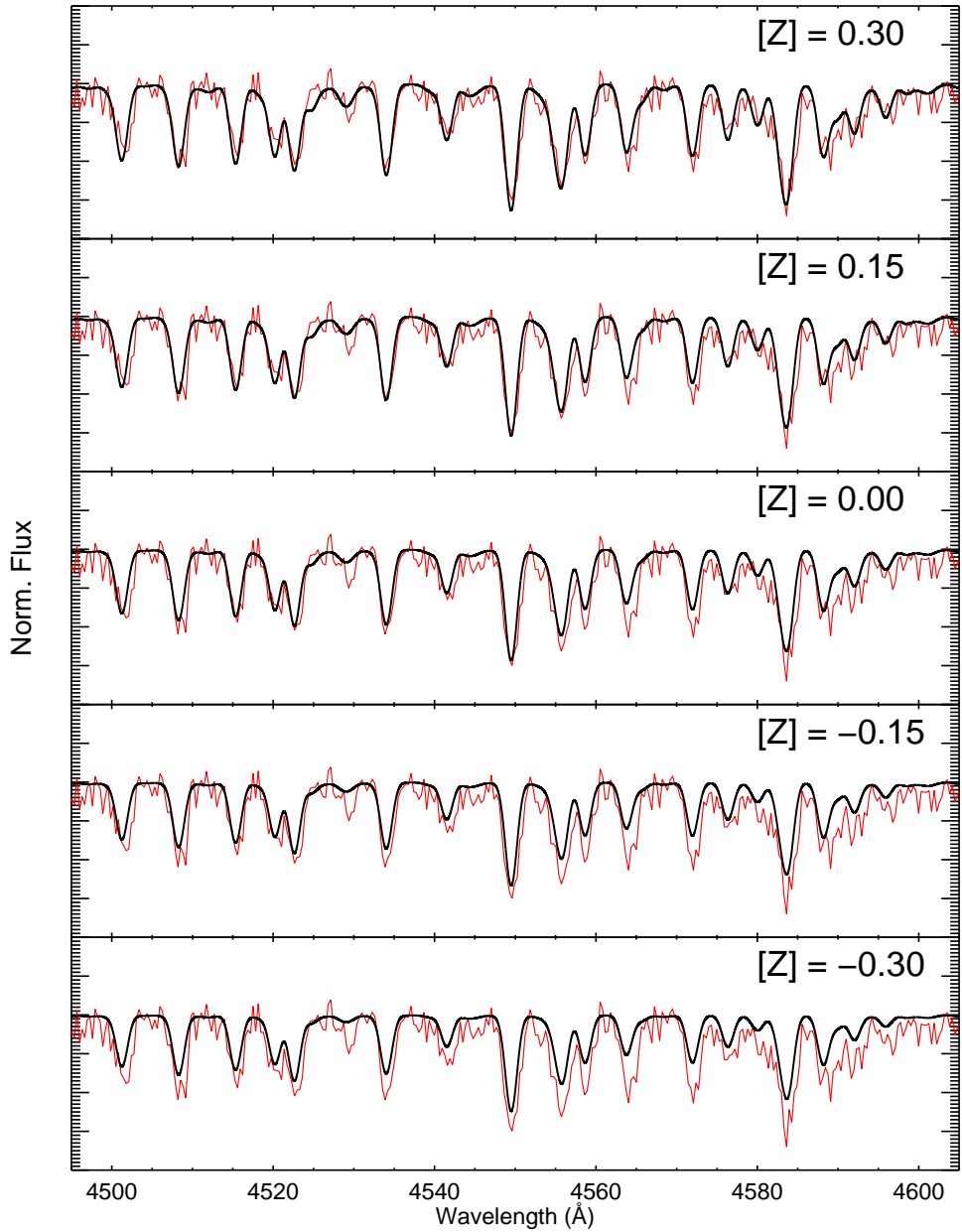}
    \caption{These panels show the metal line fit 
      for the object No. 5.  The observed spectrum (thin) is overplotted
      with the model (thick) with $T_{\rm eff}$ = 8750 K and $\log~g$ = 1.30
      dex at increasing metallicities.  The best-fit metallicity
      is $\log~(Z/Z_{\sun})$ = 0.15 (second row).}
    \label{fig:metalfit}
  \end{figure}

The use of ionization equilibrium such as Mg~{\sc i/ii} as illustrated in 
\fig \ref{fig:tfit_new_1} presumes a stellar magnesium 
abundance or, at least, a stellar metallicity (the dependence of the Balmer 
or Helium lines on metallicity is very weak and can be neglected). As a starting 
point for $T_{\rm eff}$ and $\log\,g$ determination, we have adopted metallicities 
in agreement with the metallicities and the metallicity gradient found by 
\cite{Urbaneja05}. After constraints have been placed on the values of
$T_{\rm eff}$ and $\log\,g$, we then check whether the initial
metallicity assumption was correct. This process is exemplified in
\fig \ref{fig:metalfit} and detailed as follows.  We define spectral windows, 
which are dominated by metal lines and for which a good definition of the continuum 
is possible. We then select all available models with different metallicities (typically 
ranging from [$Z$] = -1.3 to 0.3 dex, with [$Z$] = $\log\,Z/Z_{\odot}$) at the 
corresponding ($T_{\rm eff}$, $\log\,g$) and, for each of these
models, carry out a pixel-by-pixel comparison of the calculated and  
normalized fluxes within each spectral window. The metallicity over
each spectral range is thus the metallicity associated with the
minimum $\chi^{2}$-value; the average [$Z$] from all windows then gives an estimate 
of the stellar metallicity. If the value of [$Z$] thus obtained differs 
from that used for the ionization equilibrium by more than 0.1 dex, we reiterate
the $T_{\rm eff}$ and $\log\,g$ determination process using this new
[$Z$] until the metallicity measurements converge. This procedure is repeated
for models at the extremes of the error box for $T_{\rm eff}$ and $\log\,g$, 
which, together with the dispersion in [$Z$] obtained in all spectral windows, defines 
the uncertainty in [$Z$] (see K08 for details). The stellar 
metallicities with uncertainties for our sample of supergiants are given in
\tbl \ref{tbl:summary}. Note that the metallicities obtained via this technique for the late B 
and early A supergiants reflect mostly the abundances of heavy elements such as iron and 
titanium. For the early B supergiants, the same method as was
described in \cite{Urbaneja05} was applied: the metallicities are an
average of oxygen, magnesium and silicon abundances.

 \subsection{Interstellar Reddening, De-reddened Magnitudes, Stellar Luminosities, Radii, and Masses}
With effective temperature, gravity, and metallicity constrained from our 
spectral analysis, we know the intrinsic energy distribution of our target 
stars from the flux distribution of the model atmosphere computed with the 
final stellar parameters~\cite[see][]{Przybilla06,Kudritzki08,Urbaneja08}. 
Comparing with the observed fluxes from
\cite{Massey06} and applying the interstellar extinction law found by 
\cite{Cardelli89}, we can obtain an accurate 
estimate of interstellar reddening $E(B-V)$ for each individual supergiant
(see K08 for details). Assuming $A_V\,=\,3.1\,E(B-V)$ for the 
relationship between reddening and visual extinction yields the de-reddened 
apparent bolometric magnitude $m_{\rm bol}$ given by
  \begin{equation}
    m_{\rm bol}\,=\,m_V\,-\,A_V\,+BC~,
  \end{equation}
where $m_V$ is the observed apparent visual magnitude; $BC$ is the bolometric 
correction, which is also given by the flux distribution of the model atmosphere 
calculated with the final stellar parameters. Note that K08  
give an accurate analytical formula to compute $BC$ as a function of 
$T_{\rm eff}$, $\log\,g$, and [$Z$]. Stellar photometry including
intrinsic colors, reddening, and bolometric correction is summarized 
in~\tbl \ref{tbl:photo}.

Assuming an appropriate distance modulus, for instance as determined via the 
FGLR method (see below), the de-reddened apparent bolometric magnitude 
$m_{\rm bol}$ can subsequently be used to evaluate the absolute bolometric magnitude 
$M_{\rm bol}$ or, equivalently, the stellar luminosity $\log\left(L/L_{\odot}\right)$. Stellar 
luminosities coupled with effective temperatures then yield stellar radii.
\tbl \ref{tbl:masses} lists the absolute bolometric magnitudes and radii obtained in 
this way.

As explained by K08, there are two ways to determine stellar 
masses. We can use stellar gravities together with the radii to directly calculate 
masses; masses subsequently determined are referred to as spectroscopic masses. 
Alternatively, masses can be estimated by comparing the location of our target 
stars on the H-R diagram with evolutionary
tracks~\cite[][]{Maeder05}; these masses are called evolutionary
masses. Both masses along
with their uncertainties are given in~\tbl \ref{tbl:masses}.

 \section{Results}
 \label{Results}

In this section we discuss the main results of our quantitative
spectral analysis that are 
compiled in~\tbl \ref{tbl:summary},~\tbl \ref{tbl:photo}, and~\tbl \ref{tbl:masses}.

  \subsection{Interstellar Reddening and Extinction}
  \label{Reddening}

\begin{figure}
    \centering
    \includegraphics[angle=90,width=1.1\textwidth]{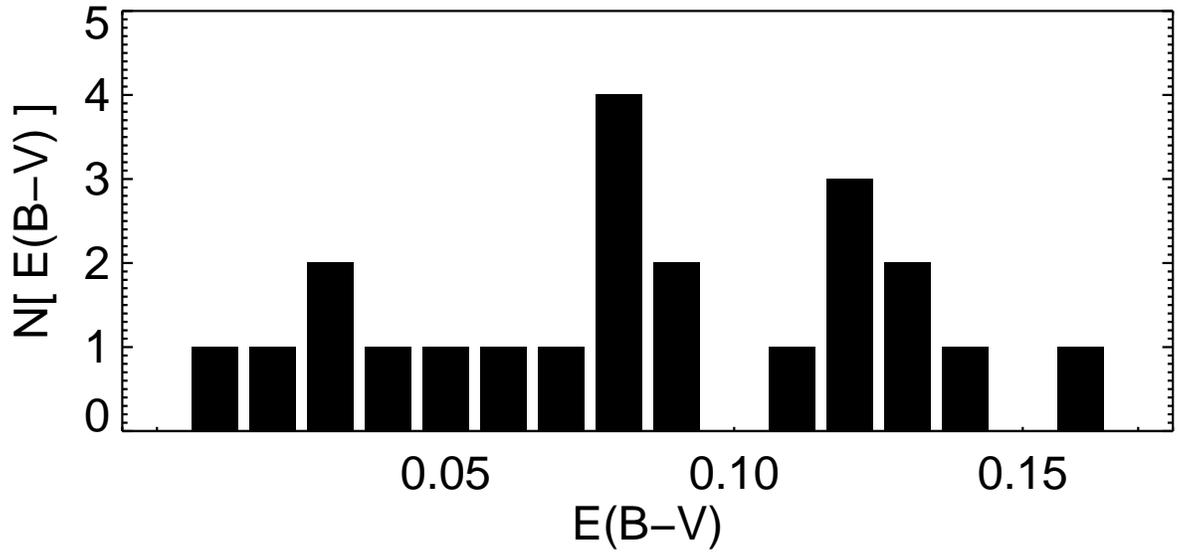}    
    \caption{The distribution of reddening $E(B-V)$ for our B and A
	supergiants; the average $<E(B-V)>$ for our sample is 0.083 mag.}
    \label{fig:reddening_new}
  \end{figure}

\fig \ref{fig:reddening_new} shows the distribution of interstellar reddening 
$E(B-V)$ among the stars in our sample. We find a wide range from 0.01 to 0.16 
mag with an average value of $<E(B-V)>$\,=\,0.083\,mag. The individual
values are significantly lower than $E(B-V)$\,=\,0.20\,mag, the reddening value as adopted in 
the HST distance scale Key Project study of Cepheids 
by~\citet{Freedman91} \citep[see also][]{Lee02}. This will have important repercussions for 
the discussion of the distance to M\,33 in \S \ref{Discussion}. We note
that our average value is higher 
than the foreground value of $E(B-V)$\,=\,0.04\,mag, which has been derived from the 
reddening maps by~\cite{Schlegel98}. We also note that 
the wide range in reddening found in this study is similar to the range in 
NGC\,300 derived by K08 and demonstrates the need for reliable 
individual reddening determinations.

\begin{figure}
    \centering
    \includegraphics[]{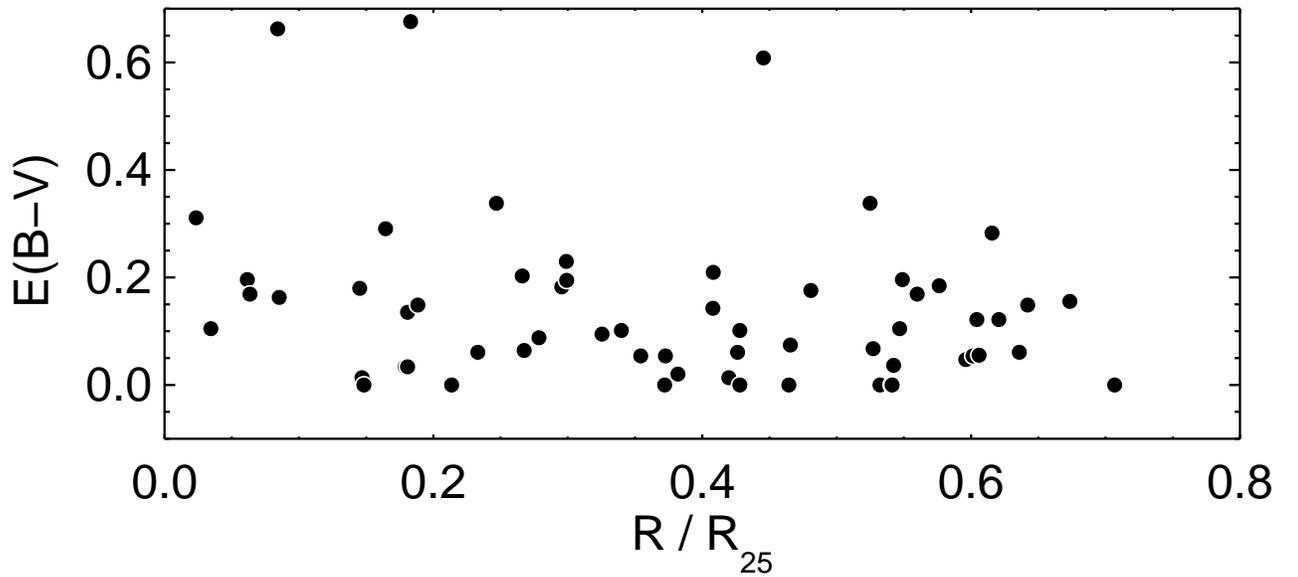}  
    \caption{Reddening $E(B-V)$ of M\,33 \ion{H}{2} regions as a function of
      angular galactic distance~\cite[$R_{25}\,=\,35.40^\prime$;][]{deVaucouleurs95}.  
      The data have been taken from \cite{Rosolowsky08}, see text. 
      The average $<E(B-V)>$ for this sample is 0.11 mag (not including the three 
      extreme objects with reddening larger than 0.6 mag).}
    \label{fig:reddening_roso}
  \end{figure}

An independent method to acquire information about interstellar reddening in 
M33 is the study of the Balmer decrement of \ion{H}{2} regions. While
\ion{H}{2} regions and the associated stars within are generally
younger than A and B supergiants as well as
Cepheids, which in many cases have already migrated into the field, and while 
the average reddening of \ion{H}{2} regions is hence very likely somewhat higher, 
it is still useful to discuss $E(B-V)$ values obtained with this
technique. Most recently, \cite{Rosolowsky08} in their 
comprehensive study of M\,33 \ion{H}{2} regions have published reddening values 
c(H$_{\beta}$) for a large sample. Adopting $E(B-V)\,=\,$0.676$\,$c(H$_{\beta}$) 
\cite[][]{Magrini07b}, we produce \fig \ref{fig:reddening_roso} with
reddening values from \cite{Rosolowsky08}, where we plot $E(B-V)$ 
as a function of angular galactic distance, in units of $R_{25}$ 
\cite[][]{deVaucouleurs95}. The result is very
informative: there is a large scatter in reddening at all distances;
however, the average $E(B-V)$ value (excluding the three most extreme
cases with $E(B-V)\,\ge\,$0.6 mag) is $<E(B-V)>$\,=\,0.11\,mag, which is in close
agreement to what has been found for our blue supergiants. We note
that \cite{Magrini07b} quote an average reddening of   
$<E(B-V)>$\,=\,0.22\,mag from their study of \ion{H}{2} regions; however, their work systematically 
overestimates the reddening value, since, when measuring Balmer emission line 
fluxes and the Balmer decrement, they have not corrected for the effect of 
the underlying stellar Balmer absorption.

  \subsection{Stellar Properties}
  \label{Properties}

\begin{figure}
    \centering
    \subfigure{\label{fig:lgt}
      \includegraphics[angle=90,width=0.8\textwidth]{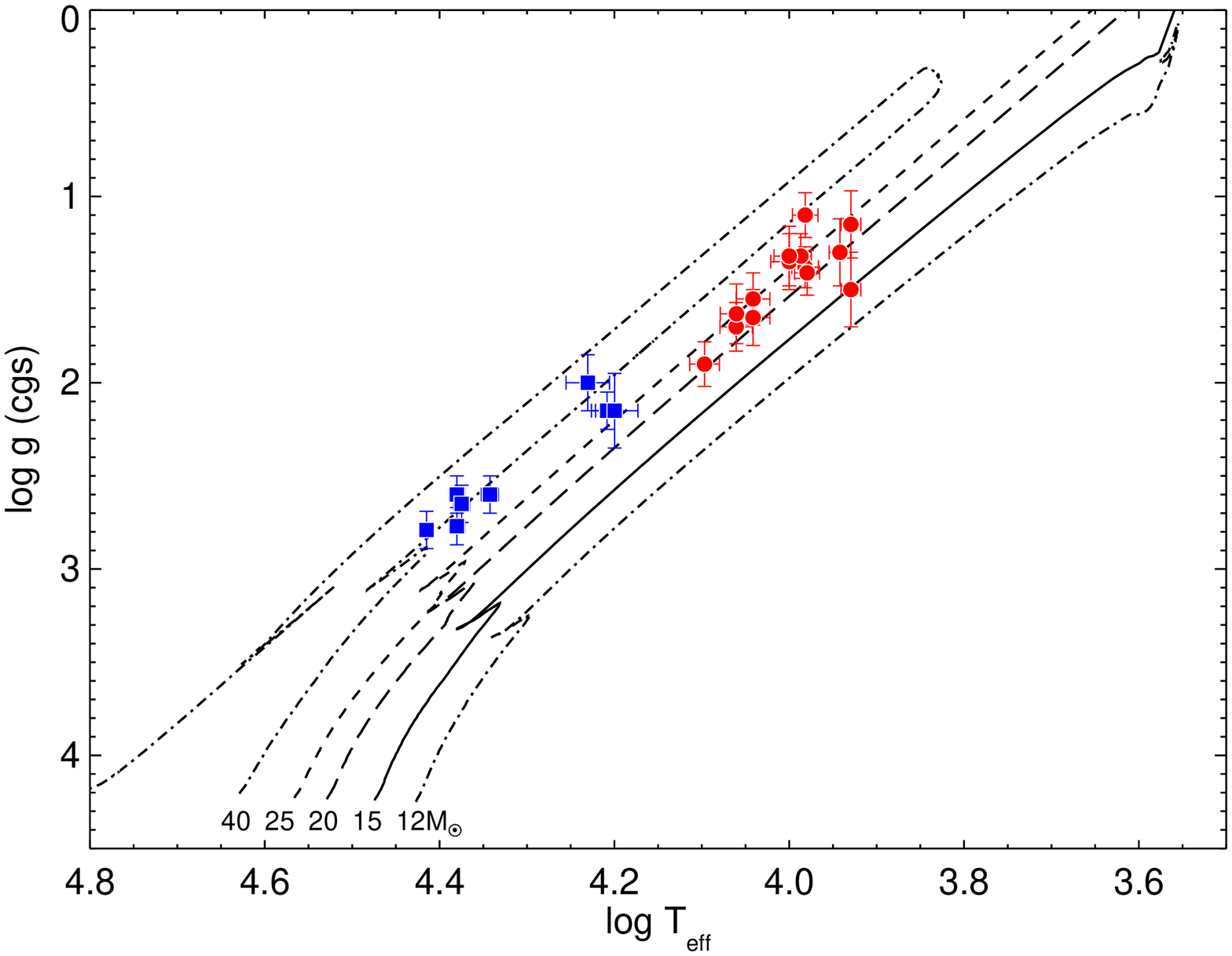}}
    \hspace{.1in}
    \subfigure{\label{fig:hrd}
      \includegraphics[angle=90,width=0.8\textwidth]{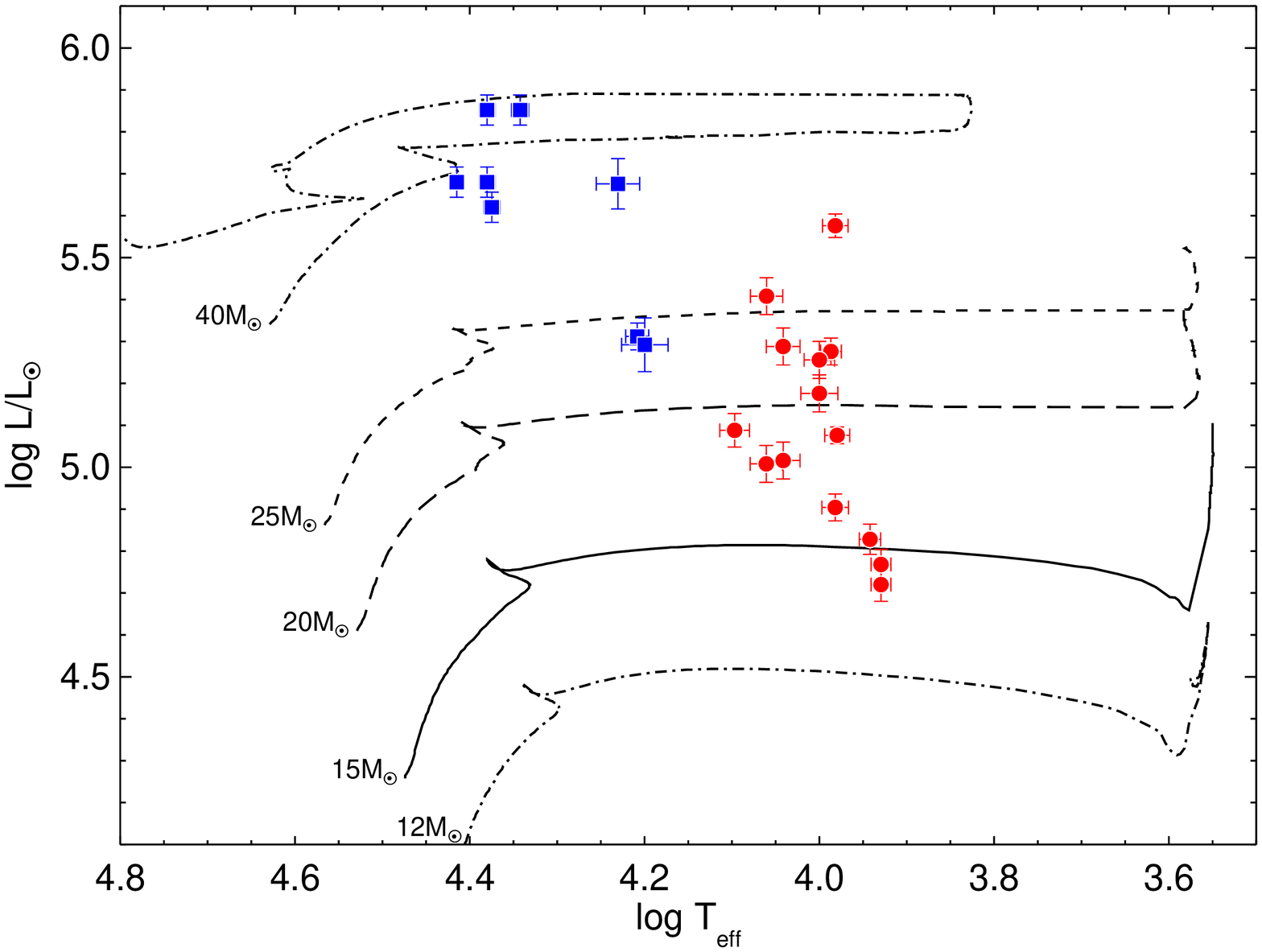}}      
    \caption{The M\,33 supergiants of this study in the
      $\log\,g$--$\log\,T_{\rm eff}$ plane (top) and in the 
      H-R diagram (bottom). Early B types are shown as solid squares,
      late B and A types as solid circles. Evolutionary tracks
      \cite[][]{Meynet03} for different ZAMS masses are overplotted.  
      \label{fig:m33_lgt}
    }    
  \end{figure}

\fig \ref{fig:m33_lgt} (upper panel) shows the location of all targets on the 
($\log\,g$, $\log\,T_{\rm eff}$)-plane along with evolutionary tracks 
\cite[][]{Meynet03} calculated for solar metallicity that include
effects of rotation and anisotropic mass loss. In this diagram, the
data points of the stars stand independent of any assumption 
about the distance to the galaxy and rely completely on 
quantitative stellar spectroscopy. There is an indication of two evolutionary 
sequences. The six early B supergiants (No. 17 to 22 on our target list) and one
late B supergiant (No. 16) align with the 40 $M_{\odot}$ evolutionary track, while 
the others appear to be in the 20 to 25 $M_{\odot}$ range. One object (No. 4) 
that coincides with the 15 $M_{\odot}$ evolutionary track  seems to be of lower mass.
The fact that all the early B supergiants are more massive is a selection 
effect: the objects have been chosen according to their apparent visual 
magnitudes. Since the early B supergiants are hotter than the rest of the sample, 
their bolometric corrections and hence their luminosities and,
consequently, their masses are larger (see discussion in K08). 

Complementary information about the evolutionary status of the targets can be 
derived from their location on the H-R diagram. With a distance modulus of 
$\mu$\,=\,24.93$\,\pm\,$0.11\,mag as determined in \S \ref{Distance}, we have calculated stellar radii and 
luminosities as described in the previous section. This allows us to position all 
targets on the H-R diagram in another comparison with the same evolutionary tracks 
(see \fig \ref{fig:m33_lgt}, bottom panel). The advantage of using this diagram is that the 
theoretical dependence of stellar luminosity on mass is very strong and, 
given the size of the error bars, allows for a more accurate assessment of the
original stellar masses. Most obviously, the same objects that appear to be 
more massive in the previous plot are also the more luminous ones associated 
with higher-mass evolutionary tracks. This nicely confirms the purely spectroscopic 
analysis, at least qualitatively.

However, we encounter a problem at lower luminosities. Now there are three objects 
(instead of just one) located on the 15 $M_{\odot}$ track at lower effective 
temperature. They are No. 2, 4, and 5 in order from lower to higher luminosity. 
Target No. 2 is particularly suspicious: it has the lowest luminosity of all 
targets indicating an original mass of less than 15 $M_{\odot}$, but on the
($\log\,g, \log\,T_{\rm eff}$)-plane it has the lowest gravity and 
effective temperature, resembling an object of significantly higher mass.
This indicates a discrepancy between the spectroscopic result and stellar 
evolutionary path for this specific object. 

\begin{figure}
    \centering
    \includegraphics[]{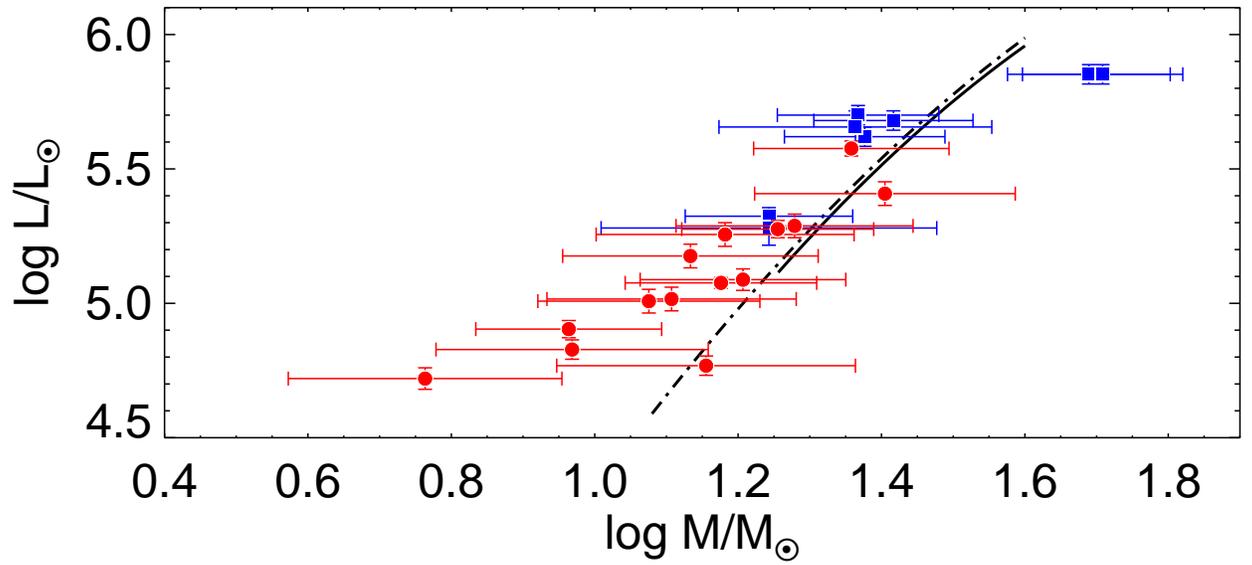}    
    \caption{The observed mass-luminosity relationship compared with the 
    relationships (dashed: late B and A supergiants, solid: early B
    supergiants) obtained from the evolutionary tracks of \fig
    \ref{fig:m33_mass_1}. Spectroscopic masses derived from stellar
    gravity and radius are used for the M 33 targets of this study.  
  }
  \label{fig:m33_mass_1}
 \end{figure}

The discrepancy becomes more apparent when we plot 
spectroscopic stellar masses against stellar luminosities and compare
the results with the 
corresponding relationship derived from currently available stellar evolutionary 
tracks at the effective temperatures of early B supergiants and A supergiants 
(see K08 for a simple fit formula). This is done in 
\fig \ref{fig:m33_mass_1}, which reveals that the spectroscopic mass of target 
No. 2 (the object with the lowest mass and luminosity) does not fit 
on the stellar evolutionary mass--luminosity relationship. Its mass is clearly too 
low for its present luminosity. We note that K08 in their study 
of blue supergiants in NGC\,300 have found a very similar outlier object 
(their target No. 17). They speculated that these objects could be stars 
of lower initial mass now evolving back from the red giant branch to
become blue supergiants again at significantly higher luminosity.
However, an inspection of the tracks by 
\cite{Meynet03} does not support this speculation. None of the tracks 
with 12, 15, and 20 $M_{\odot}$ loops back from the red giant branch, whereas the 
tracks with significantly lower masses do not reach the luminosity of object 
No. 2 within their paths. A more attractive explanation appears to be close binary 
evolution during which No. 2 (and No. 17 of K08) might have 
lost a significant fraction of their original mass while roughly keeping their 
original luminosity. Expectedly, such a scenario will need to be confirmed by 
detailed evolutionary calculations. Regardless of the details, it seems that this 
object has not followed the same evolutionary path as the rest of our
sample. It is obvious that the star will not fit into the FGLR
defined by {\it regular} supergiant stars. Therefore, for distance
determination using the FGLR-method we will exclude
object No. 2, in the same way K08 excluded their object No. 17.

\begin{figure}
    \centering
     \subfigure{\label{fig:mass_m33}
      \includegraphics[]{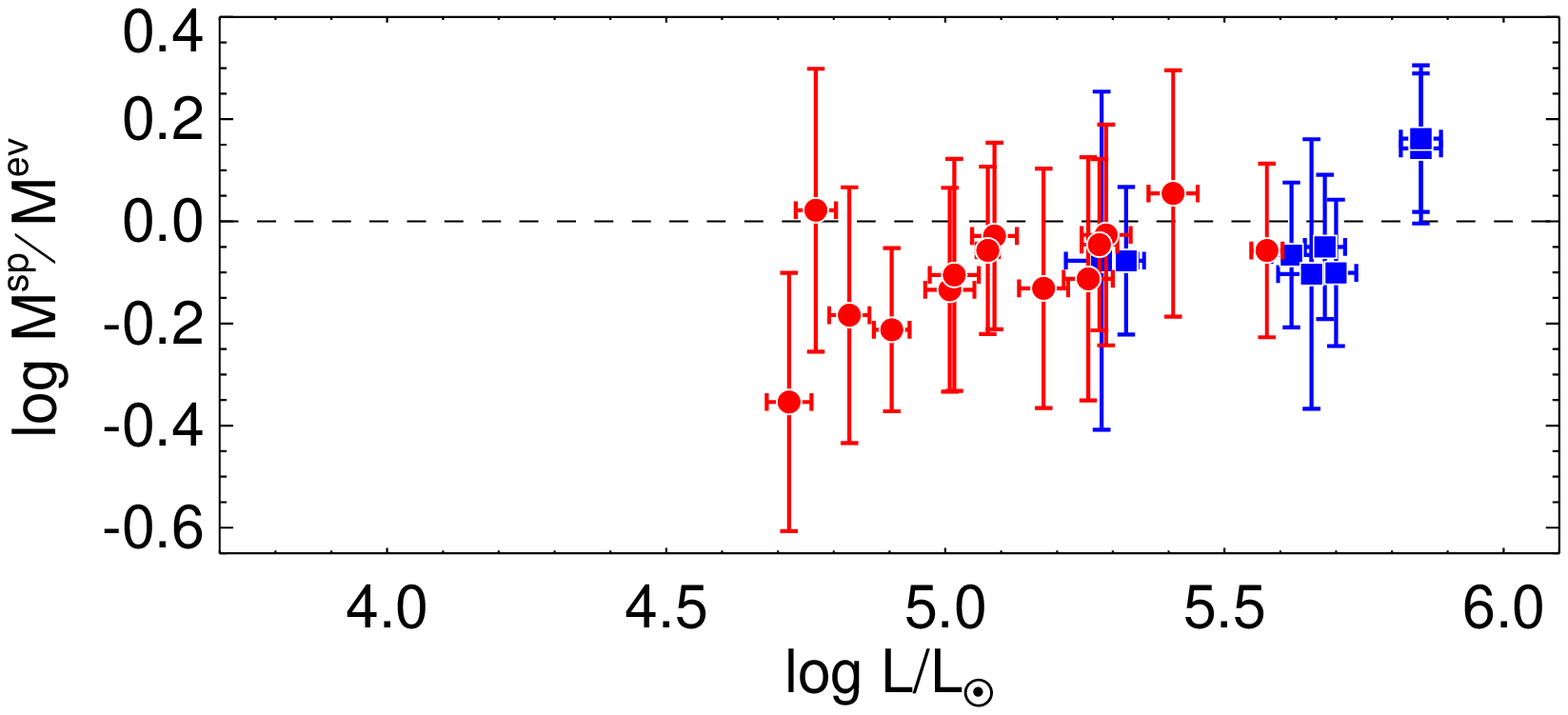}}
   \hspace{.1in}
     \subfigure{\label{fig:mass_all}
       \includegraphics[]{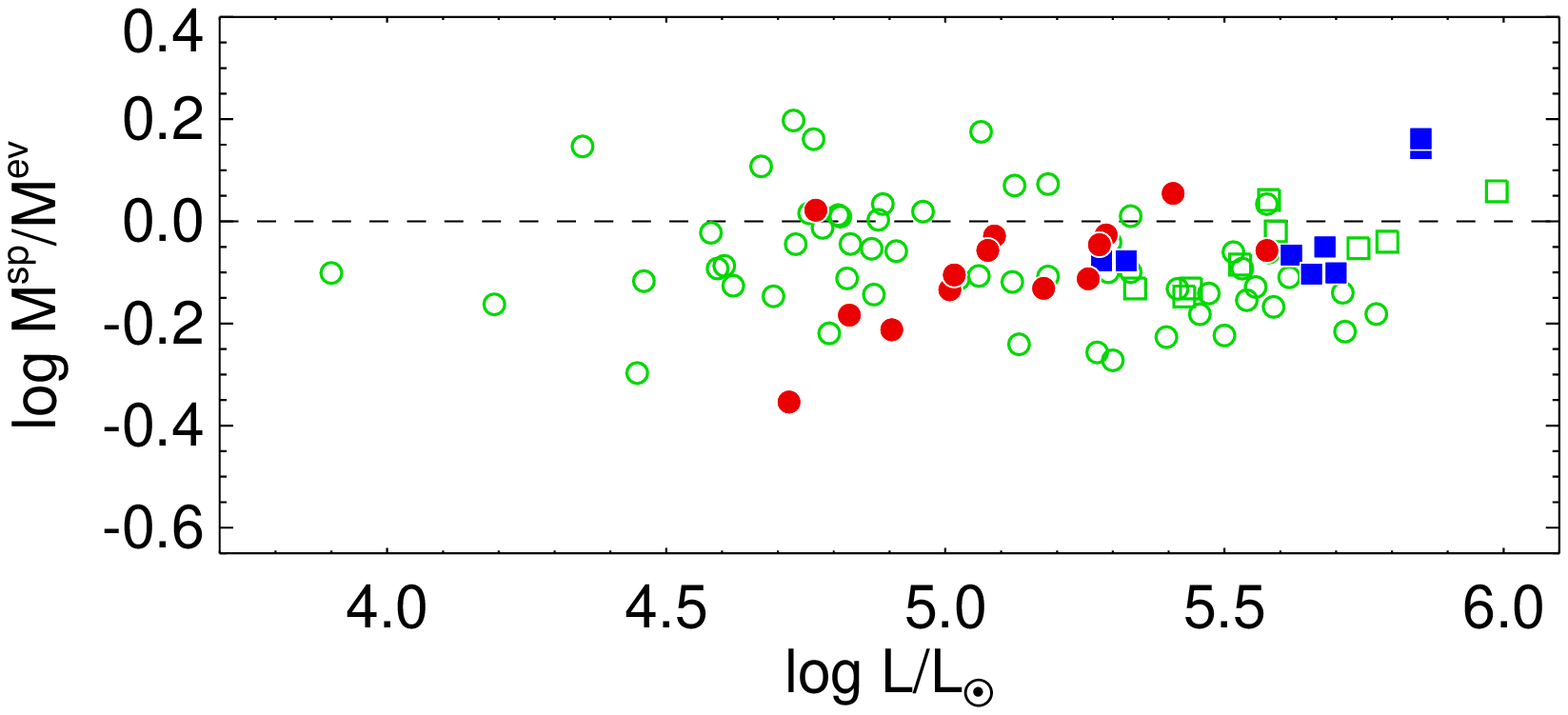}}
   \caption{Logarithmic ratio of spectroscopic to evolutionary masses
     as a function of luminosity. The upper diagram contains
     only the M\,33 objects of this study, whereas the lower
     diagram includes the results by K08 and \cite{Urbaneja08}.}
    \label{fig:m33_mass_2}
  \end{figure}

\fig \ref{fig:m33_mass_1} seems to indicate that the observed relationship 
between spectroscopic masses and luminosities for the remaining objects is in 
slight disagreement with stellar evolution models since most of the 
objects are somewhat overluminous for their mass. In order to discuss this more 
quantitatively we use the stellar evolutionary mass--luminosity 
relationships for A and B supergiants in \fig \ref{fig:m33_mass_1} (see also 
the fit formulae in K08) to convert observed stellar luminosities 
into evolutionary masses (as described in \S \ref{Spectral Analysis}). 
In  \fig \ref{fig:m33_mass_2} we plot the logarithmic ratio of 
spectroscopic to evolutionary masses as a function of luminosity. As we can see 
from these plots and \tbl \ref{tbl:masses}, spectroscopic masses are
on average 0.06 dex lower than their evolutionary counterparts (not
including No. 2 in calculating the average), indicating indeed 
a small systematic discrepancy between spectroscopic results and
stellar evolution theory. We note that while evolutionary masses have
much smaller error bars than spectroscopic masses do due to the steep
dependence of luminosity on mass,
they might be subjected to systematic uncertainties such as the influence of 
mass loss over the stellar lifetime or rotationally-induced mixing effects
on theoretical luminosities. Of course, spectroscopic masses might 
be subjected to systematic uncertainties as well, in particular through the way
of pressure broadening of the Balmer lines is being accounted for in
the process of spectroscopically determining stellar gravities as
demonstrated in the previous sections.

In order to assess whether or not the discrepancy encountered is specific to our 
M33 sample, we repeat the logarithmic ratio of the spectroscopic to
evolutionary masses analysis for stars in other galaxies in the bottom
panel of \fig \ref{fig:m33_mass_2}. This time we examine targets from 
the eight galaxies used  for the FGLR calibration by K08 and from the most 
recent study of WLM by \cite{Urbaneja08}. We do not see any indication of a special  
systematic effect that is only related to M\,33. The average logarithmic ratio for the 
66 targets overplotted is -0.04\,dex. 

Due to the fact that both evolutionary and spectroscopic masses depend on
a presumed luminosity, it is worthwhile to discuss whether or not
an inappropriate choice of the distance modulus would cause a
systematic effect. Since $M_{\rm spec} \propto gR^{2} \propto d^{2}$, $L \propto d^{2}$, 
and a mass-luminosity relationship of the form 
$L \propto M_{\rm evol}^{\alpha}$, we derive a logarithmic ratio of 
spectroscopic to evolutionary masses with dependence on $\alpha$ and
$d$ as the following: 

 \begin{equation}
    \log M_{\mathrm{spec}}/M_{\mathrm{evol}}\,=\,(2-2/\alpha)\log\,d
    \quad .
  \end{equation}

With $\alpha \sim 3.5$ (see \fig \ref{fig:m33_mass_1} or K08), 
we obtain $\log M_{\rm spec}/M_{\rm evol}=1.4\log\,d$, which means that the distance 
modulus for M33 would have to be 0.2 mag larger than what we have
adopted.  As the discussion of the distance to M33 in \S
\ref{Distance} will show, this is unlikely.

\subsection{Metallicity and Metallicity Gradient}
 \label{Metallicity}

\begin{figure}
    \centering
    \includegraphics[width=0.9\textwidth]{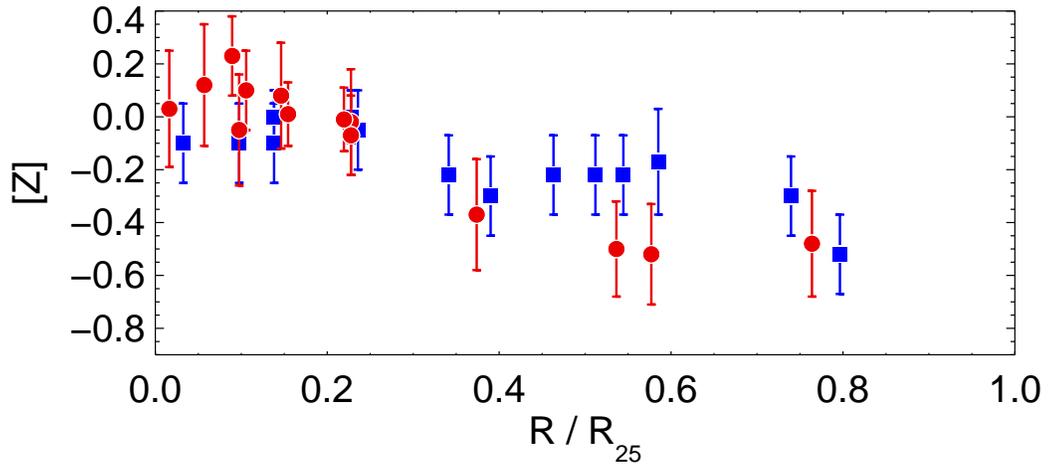}
    \caption{Blue supergiant metallicity [$Z$] as a function of
      dimensionless angular galactocentric distance. Circles:
      late B and A supergiants; squares: early B
      supergiants.  \label{fig:met_grad} } 
  \end{figure}

Findings from our quantitative spectroscopic study can be 
used to assess the metallicity of the young stellar populations in M33 
as well as the results of emission line studies of \ion{H}{2} regions. 
\fig \ref{fig:met_grad} displays the stellar metallicities obtained in
consequence as a function of dimensionless angular galactocentric
distance. In this, and the corresponding following plots, we use
$R_{25}$\,=\,35.40$^\prime$~\cite[][]{deVaucouleurs95}. Our basic
outcome is that the stellar metallicity is close 
to the solar value at the center and that it decreases by a factor of
three at the galactocentric distance of an isophotal radius. A simple
linear regression of the form
 \begin{equation}
    [Z]\,=\,[Z]_{0}\,+\,[Z]_{1}\left( R / R_{25} \right)
  \end{equation}
yields $[Z]_{0}\,=\,$0.09$\,\pm\,$0.04\,dex for the central metallicity and 
$[Z]_{1}\,=\,$-0.73$\,\pm\,$0.09\,dex\,$R_{25}^{-1}$ for the 
angular gradient (equivalent to -0.07$\,\pm\,$0.01
dex\,kpc$^{-1}$  for the distance used in this paper). We note,
however, that the scatter is significant around 0.5\,$R_{25}$; it is
very likely a real phenomenon rather than an artifact of the
uncertainties of our analysis.  

\begin{figure}
    \centering
    \includegraphics[width=0.9\textwidth]{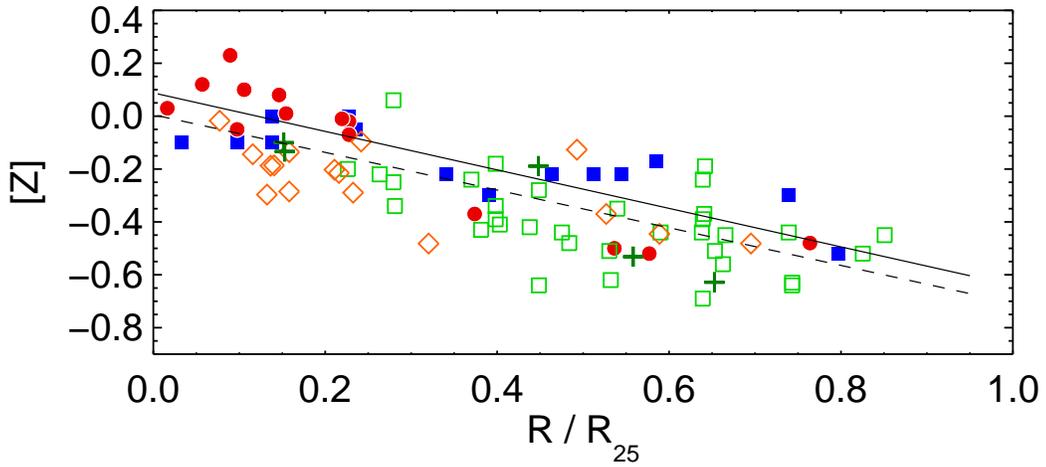}
    \caption{Metallicity of blue supergiants, \ion{H}{2} regions and
      Cepheids as a function of dimensionless angular galactocentric
      distance. A and B supergiants have the same symbols as \fig
      \ref{fig:met_grad}. Logarithmic oxygen abundances of \ion{H}{2}
      regions in units of the solar value as published by
      \cite{Magrini07a} are plotted as open squares. Logarithmic neon
      abundances of \ion{H}{2} regions normalized to the value for B
      stars in the solar neighbourhood and as obtained from
      \cite{Rubin08} are shown as large open diamonds. The metallicity
      [$Z$] for beat Cepheids as determined by \cite{Beaulieu06} are
      given as crosses. The solid line is the regression for the
      supergiants only, whereas the dashed lines is the regression for
      all objects. 
}
    \label{fig:met_grad_magr}
  \end{figure}

\fig \ref{fig:met_grad_magr} is similar to \fig \ref{fig:met_grad}, 
but it also includes the \ion{H}{2} region Oxygen abundances compiled by 
\cite{Magrini07a}, \ion{H}{2} Neon 
abundances as determined most recently by \cite{Rubin08}, 
and Cepheid metallicities as obtained by \cite{Beaulieu06}. (We found 
that the de-projected galactocentric distances given by
\cite{Beaulieu06} contained errors that we have corrected for 
\fig \ref{fig:met_grad_magr}). The value for the solar Oxygen abundance 
adopted to normalize the \ion{H}{2} data by  \cite{Magrini07a} is 
$\log\left[N(O)/N(H)\right]\,+\,$12\,=\,8.69\,dex \cite[][]{Allende01}. From this sample, 
we considered only those \ion{H}{2} regions for which a determination of the
electron temperature was possible.  Since the value of the solar Neon abundance 
seems to be rather uncertain 
\cite[see discussion in][]{Rubin08}, we normalize Neon to the average value 
for B stars in the solar neighborhood and adopt $\log\left[N(Ne)/N(H)\right]\,+\,$12\,=\,8.08 dex 
\cite[][]{Przybilla08}.

The different data sets seem to converge: they all indicate a significant abundance 
gradient. Combining both sets of \ion{H}{2} region abundances, we derive a slope
of -\,0.55$\,\pm\,$0.09 dex\,$R_{25}^{-1}$ (-0.06$\,\pm\,$0.01 dex\,kpc$^{-1}$), statistically consistent  
with the stellar results. With respect to the intercept, a comparison
is less robust since several issues could be affecting the zero point
of the different abundance scales. The nebular data provide
-0.11$\,\pm\,$0.04 dex, slightly below the stellar result, but
nevertheless consistent and not unexpected given the systematics that
could be affecting both samples (stars and ionized gas). We regard
these as consistent results. A least-square fit to all objects in this
diagram yields [$Z$]$_{0}\,=\,$-0.01$\,\pm\,$0.03 dex for the central value and 
[$Z$]$_{1}\,=\,$-0.71$\,\pm\,$0.07 dex\,$R_{25}^{-1}$ for the gradient. 
Within the uncertainties, this agrees with the results obtained from the supergiant stars
alone. We note again that the results for the \ion{H}{2} regions displayed in this 
figure also indicate a large scatter around the regression curve.

\begin{figure}
    \centering
    \includegraphics[width=0.9\textwidth]{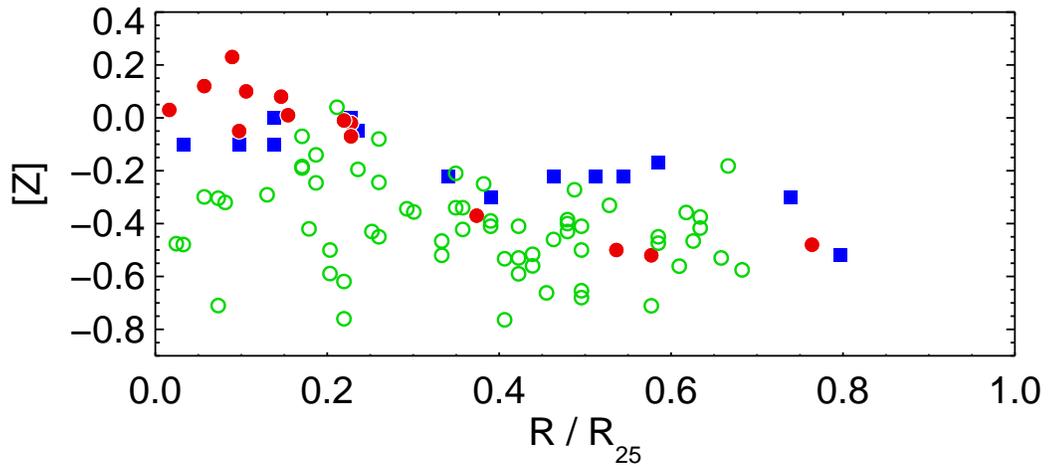}
    \caption{Blue supergiant metallicity [$Z$] as a function of
      dimensionless angular galactocentric distance compared to the
      logarithmic oxygen abundances (in units of the solar value) of
      the \ion{H}{2} regions investigated by \cite{Rosolowsky08},
      open circles.
}
    \label{fig:met_grad_roso}
  \end{figure}

This significant metallicity gradient is in obvious disagreement with the 
result found by \cite{Rosolowsky08}, who carried out a comprehensive 
study of \ion{H}{2} regions in their M33 metallicity project. \fig \ref{fig:met_grad_roso} 
compares the Oxygen abundances of our supergiants with that of their
\ion{H}{2} regions. While our data points largely coincide with theirs
at large galactocentric distances, \cite{Rosolowsky08} find many 
\ion{H}{2} regions with very low oxygen abundance in the central region of M33 for which 
we have no stellar counterparts in our sample. At this point, we have no 
explanation for this discrepancy.

\section{Distance}
 \label{Distance}

In this section we employ two different methods to determine the distance to M33. 
First, we use the stellar parameters from this spectroscopic study and
apply the FGLR method; second, we analyse Hubble Space  
Telescope (HST) Advanced Camera for Surveys (ACS) photometry of three
outer fields of M\,33 and obtain a distance from the $I$-band
magnitude of the tip of the red giant branch (TRGB method). The results from
each technique will be discussed accordingly.

  \subsection{FGLR and Distance to M33}
  \label{FGLR}

The FGLR is a tight 
correlation between the flux-weighted gravity ($g_F\,\equiv\,g/{T^4}_{\rm eff}$) 
and the absolute magnitude of BA supergiants. The physical background, detection 
and the calibration in nearby galaxies has been described 
in~\cite{Kudritzki03} and K08. ~\cite{Urbaneja08} were the first to use 
the FGLR for distance determination and found a distance modulus of 
24.99 mag for the metal-poor dwarf galaxy WLM, in good agreement with most 
recent TRGB distance determinations. Here, we follow the same
procedure as was detailed in~\cite{Urbaneja08}.

\begin{figure}
    \centering
     \subfigure{\label{fig:fglr_m33}
      \includegraphics[width=0.8\textwidth]{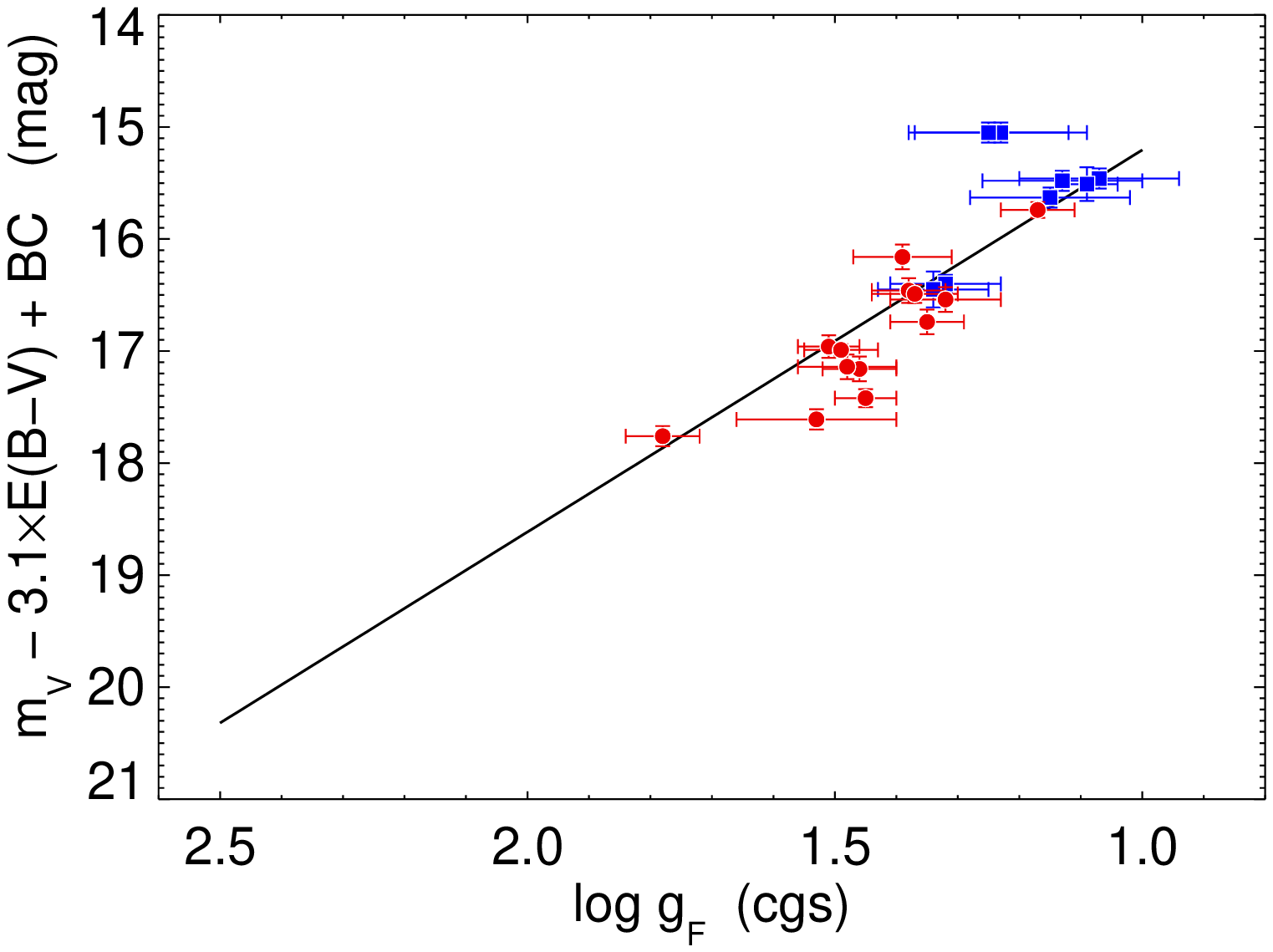}}
   \hspace{.1in}
     \subfigure{\label{fig:fglr_all}
      \includegraphics[width=0.8\textwidth]{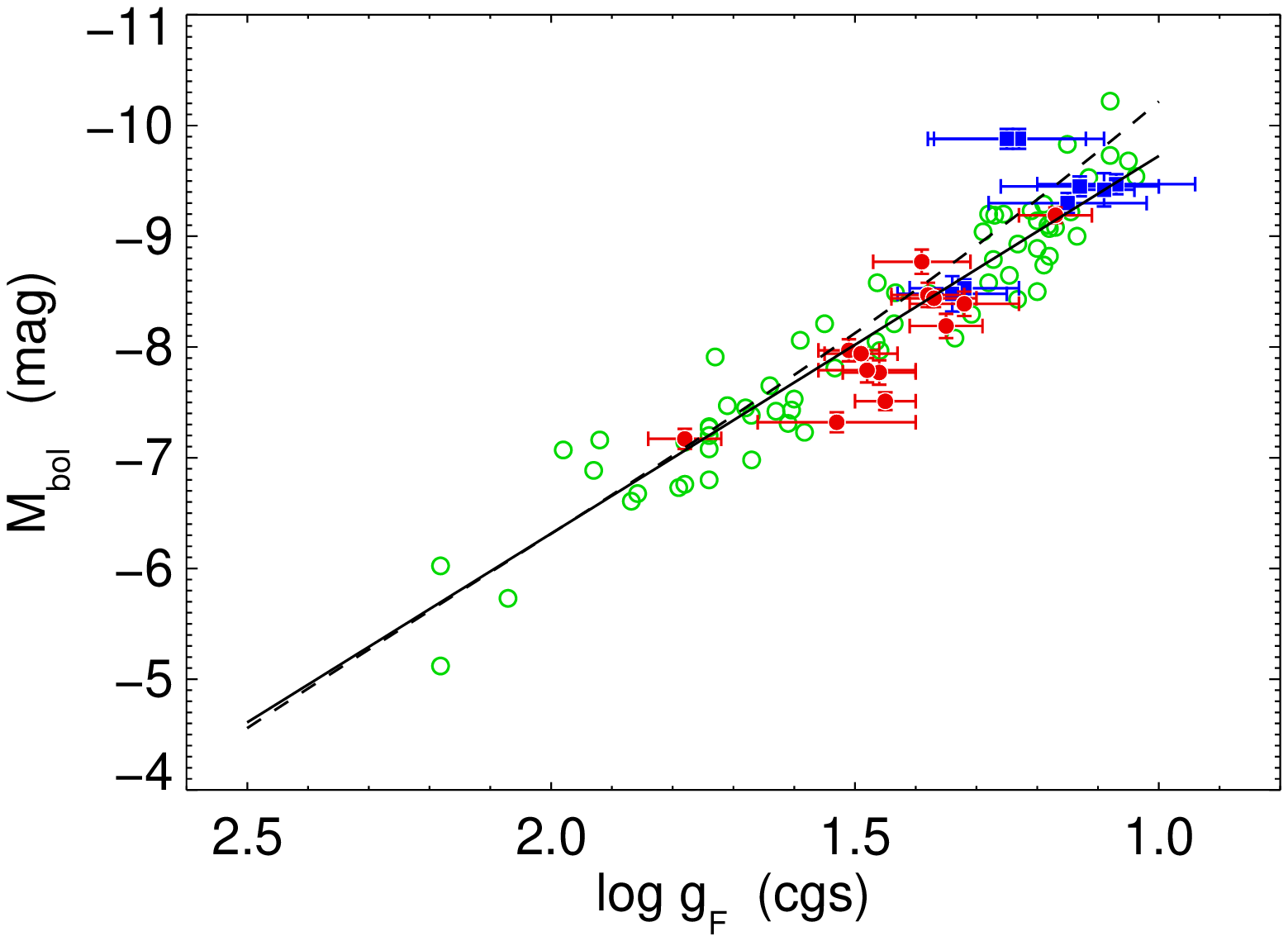}}     
    \caption{FGLR fits of the blue supergiants in M33. Top: Apparent
      de-reddened bolometric magnitude vs. flux-weighted gravity. Solid
      circles are late B and A supergiants and solid squares are early
      B supergiants in M\,33. The solid line is a linear fit as
      described in the text.  Bottom: Absolute bolometric magnitude
      vs. flux-weighted gravity. In addition to the M\,33 targets,
      objects from nine other galaxies investigated in the studies by
      K08 and \cite{Urbaneja08} are also shown. The solid line is the
      regression FGLR from K08. The dashed curve is the stellar
      evolution FGLR for Milky Way metallicity. 
   }
    \label{fig:fglr}
  \end{figure}

The FGLR has the form
\begin{equation}
    M_{\rm bol}\,=\,a (\log\,g_F\,-\,1.5)\,+\,b
  \end{equation}
with the most recent calibration provided by K08, $a$ = 3.41 and $b$ = -8.02.
Our spectroscopic analysis provides de-reddened apparent bolometric magnitude $m_{\rm bol}$
(see calculation details in \S 3.4) and flux-weighted gravity  
for each of our targets (except for object No. 2, as explained in \S
4.2) that we can fit with a regression of the form
 \begin{equation}
    m_{\rm bol}\,=\,a_{\rm M33} (\log\,g_F\,-\,1.5)\,+\,b_{\rm M33}~.
  \end{equation}
The fit result is shown in \fig \ref{fig:fglr} (top). Since our M33 targets span only 
a limited range in $g_{F}$ compared to the K08 sample, we adopted
the slope value provided by K08 (letting $a_{\rm M33}\,=\,a$) and fit
only $b_{\rm M33}$. The difference between $b$ and  $b_{\rm M33}$
yields the distance modulus, which we determine to be
$\mu$ = 24.93$\,\pm\,$0.11 mag \cite[the error is calculated similarly
as in][]{Urbaneja08}. 
\fig \ref{fig:fglr} (bottom) then compiles the FGLR results for our M33 targets using 
this distance modulus as well as that for 9 other galaxies 
from K08 and~\cite{Urbaneja08}. The plot does not 
indicate any systematic differences between the M33 sample and the remaining sample.

We note that the present calibration of the FGLR by K08 is based on a
sample of supergiants selected from NGC 300 and seven Local Group
galaxies for which distances were adopted. A cleaner, more accurate and
systematic method to calibrate the relationship would be to use a large sample
of blue supergiants in the Large Magellanic Clouds (LMC). This work is presently under way.

  \subsection{TRGB and Distance to M33}
  \label{TRGB}

\begin{figure}
    \centering
     \subfigure{\label{fig:foot}
      \includegraphics[width=0.6\textwidth]{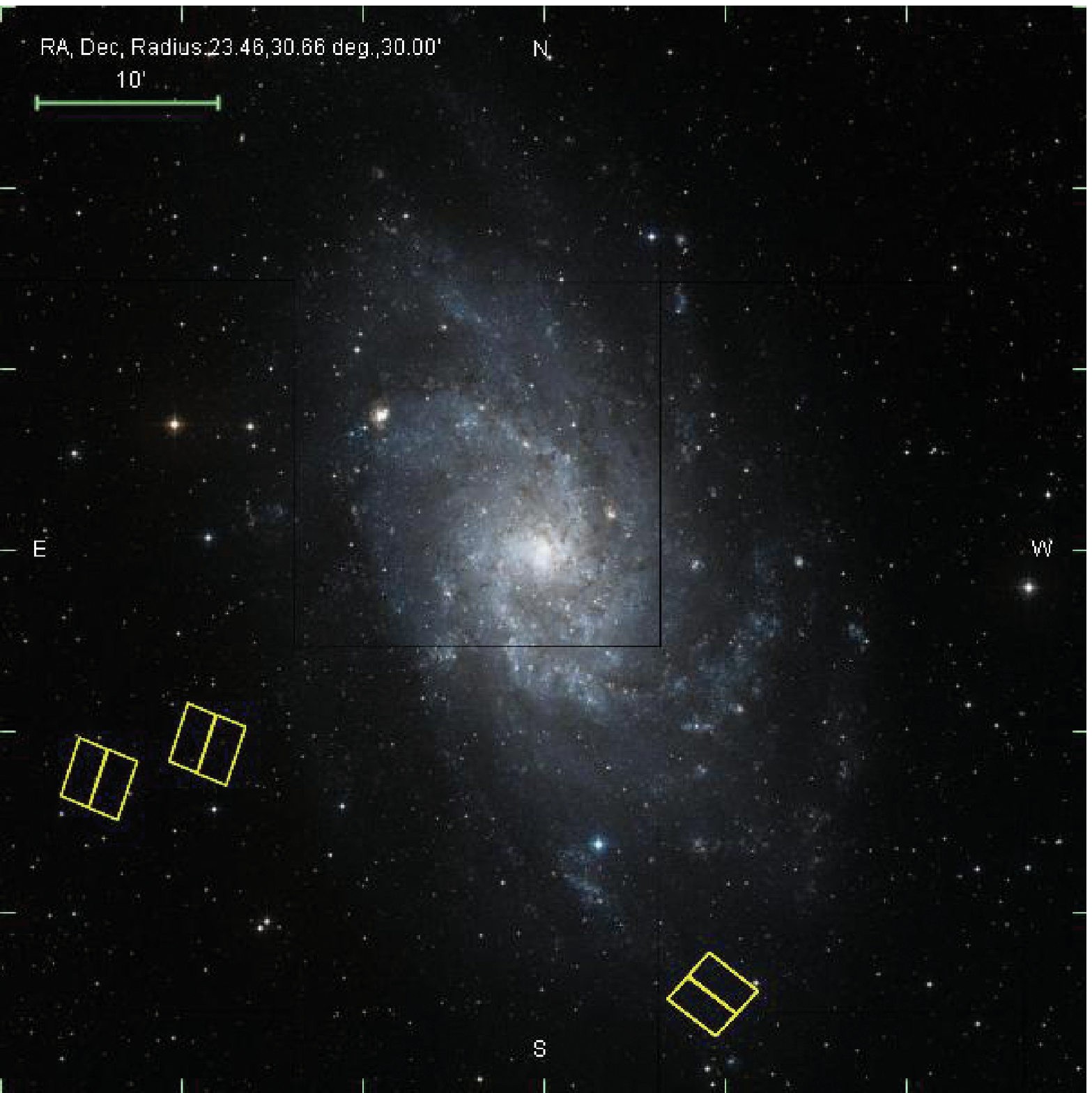}}
   \hspace{.1in}
     \subfigure{\label{fig:cmd}
      \includegraphics[width=0.6\textwidth]{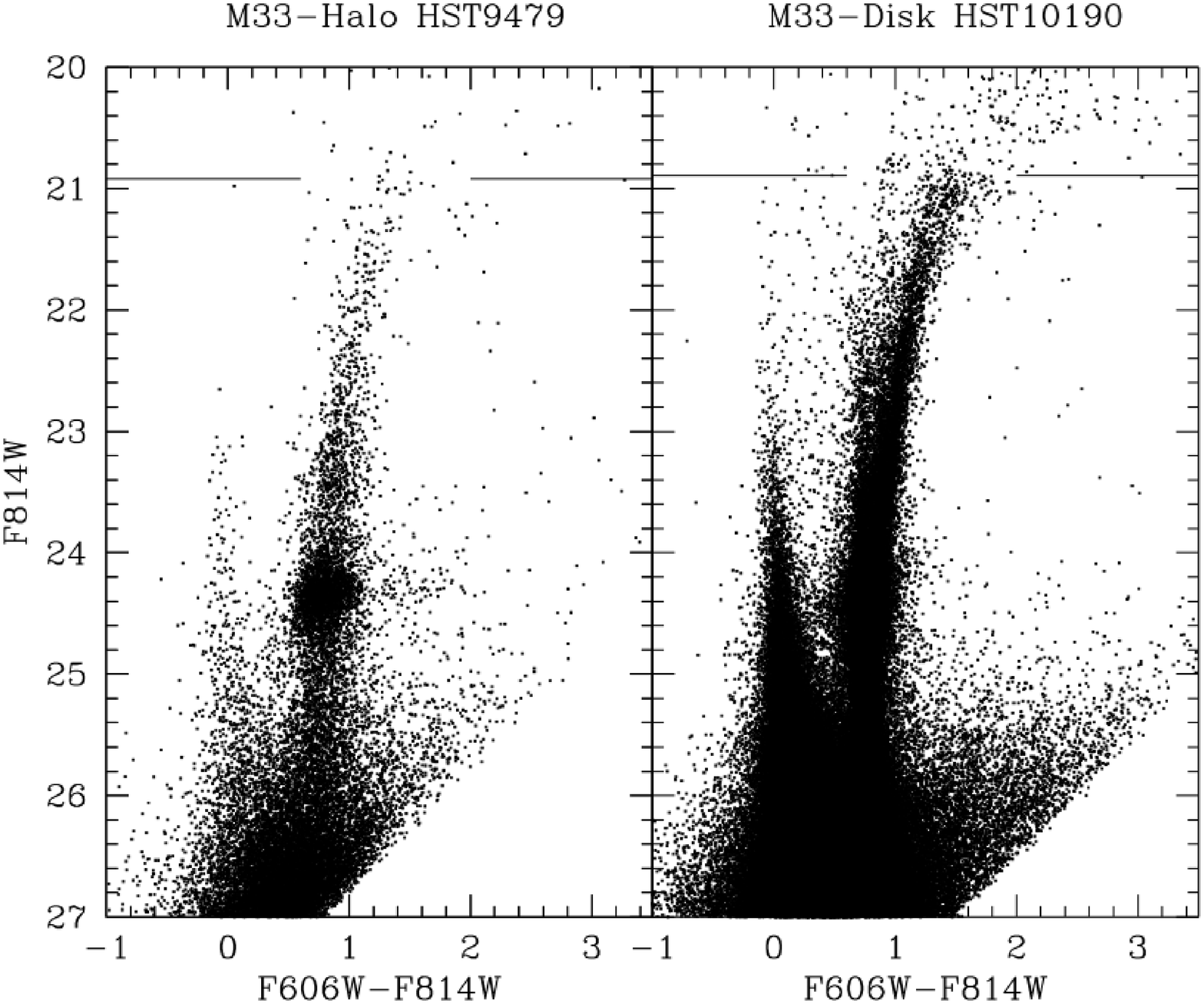}}
   \caption{Observation of the TRGB in M33. The location of the HST
     ACS halo fields and the disk field are shown in the top
     figure. The bottom figure shows the two color-magnitude diagrams
     obtained from the halo fields and the disk field. The $I$
     magnitude of tip of the red giant branch is indicated. 
}
    \label{trgb}
  \end{figure}

We use observations of M\,33 available through the HST archive to 
measure a distance based on photometry of stars at the TRGB. The footprints 
of the observations taken with ACS are  
shown in \fig \ref{trgb} (top) as yellow outlines.  The two fields  
towards the east (left) side of the image are from HST Program 9479's  
study of M33's halo; these fields are combined in the ensuing analysis.
The remaining field to the south of the image comes from HST  
Program 10190's observations of the outer disk.  We choose these sets  
of images from among the many available through the HST archive in order to  
balance the need of minimizing issues of stellar crowding and reddening  
within M33 with the need for a well-populated RGB.   

Both programs  
use the $F814W$ and $F606W$ filters, which can be described as `wide  
$I$' and `wide $V$' filters, respectively.  We perform photometry and conduct  
artificial star tests on the images using the DOLPHOT software  
package, which is a modified version of HSTphot~\cite[][]{Dolphin00}.
The resulting color-magnitude diagrams are presented in  
\fig \ref{trgb} (bottom).  Using the real and artificial photometry we  
measure the $F814W$ magnitude and color of the TRGB using a 
maximum-likelihood method described by~\cite{Makarov06}.   

The $F814W$ magnitude of the TRGB measured in the halo fields from
Program 9479 is $20.92^{+0.09}_{-0.11}$ with
$F606W-F814W=1.35\,\pm\,0.05$, while the observations of the
disk from Program 10190 yields $20.89^{+0.09}_{-0.05}$ with
$F606W-F814W=1.47^{+0.02}_{-0.08}$; the TRGB magnitudes
are marked with broken horizontal lines in \fig \ref{trgb}. The flux
from stars at the TRGB is least sensitive to age and metallicity in
the $I$-band, and has an absolute magnitude of $M_I \approx -4.05$
\cite[][]{Rizzi07}.  \cite{Rizzi07} also describe a method for
calculating a distance modulus that includes a zero-point calibration
and metallicity correction; the formula for which we reproduce below:
\begin{equation}
DM=m_{\rm TRGB}-A_{F814W}+4.06-0.20[(F606W-F814W)-(A_{F606W}-A_{F814W})-1.23]~,
\end{equation}
where $m_{\rm TRGB}$ is the apparent magnitude in $F814W$, and
$A_{F814W}$ and $A_{F606W}$ are foreground extinction for the two
filters, respectively.
We account for foreground reddening using  
the dust maps of~\cite{Schlegel98} with a value of 
$E(B-V)$ = 0.04 mag and their prescription for converting to HST
flight magnitude. Whilst we have encountered much higher 
reddening values in our study of the blue supergiants, the use of only 
foreground reddening is very likely justified in these outer fields. 
Combining these corrections with the measured magnitudes of the tip 
produces distance moduli of $\mu=24.86^{+0.09}_{-0.13}$ mag for the 
halo observations and $\mu=24.82^{+0.10}_{-0.06}$ for with the outer disk field. As 
an average TRGB distance modulus we adopt $\mu\,=\,24.84 \pm 0.10$
mag, which is in basic agreement with that found by~\cite{Rizzi07}
using HST WFPC2 data ($\mu\,=\,24.71 \pm 0.04$).

  \subsection{Discussion}
  \label{Discussion}

The two independent methods described in this paper, the FGLR and the TRGB, both yield a
large distance modulus and agree with each other within statistical uncertainties. 
The FGLR-method stands unaffected by interstellar extinction uncertainties 
because reddening has been determined for each target individually
through quantitative spectral analysis, which yields accurate stellar 
parameters and, thus, intrinsic SEDs and colors. In this regard, the FGLR 
work presented here is very similar to the quantitative spectral analysis 
of the detached eclipsing O-star binary D33 J1013346.2+304439.9 by~\cite{Bonanos06}, 
who obtained a distance modulus of $\mu=24.92\,\pm\,0.12$ mag with a 
reddening of $E(B-V)=0.09$ mag.

B and A supergiants as well as O stars belong to the young stellar population 
in the dust-obscured disk of M33. An independent and accurate 
determination of extinction hence falls inevitably from their use as distance 
indicators. The same is also true for other types of objects belonging 
to a population of similar age, such as Long Period Variables and Cepheids. 
For the former,~\cite{Pierce00} have carried out a careful 
photometric study of the period--absolute magnitude relationship of LPVs in 
Per OB1, the LMC, and M33 using $R$ and $I$
photometry and a narrow band filter at 8250\AA~(particularly to
correct for the varying strength of TiO absorption). Assuming  
a reddening value of $E(B-V)=0.1$ mag for M33, they obtained a distance 
modulus of $\mu$=24.85$\,\pm\,$0.13 mag to the galaxy. While reddening was not 
determined for individual targets, the value assumed is close to the average value 
of 0.083 mag found in our study. Thus, it seems consistent that their
distance modulus is also in agreement with ours.   

The situation with Cepheid distance determinations turns out to be more complex. 
Based on HST WFPC2 photometry from \cite{Freedman91} and \cite{Lee02},
the distance modulus was found to be $\mu=24.62$ $\pm$ 0.15 mag and  
$\mu=24.52$ $\pm$ 0.14 (random) $\pm$ 0.13 (systematic) mag, respectively. Whilst the former 
study applied a correction for metallicity, the latter did not. Both works considered a 
very high value of reddening, $E(B-V)=0.20$ mag, which was derived from the
difference in the apparent distance moduli between $V$- and
$I$-bands. This reddening value differs substantially from the average
value of 0.08\,mag found for the B and A supergiants in our study. 

Since Cepheids are also young massive stars (though less massive than
blue supergiants), they are found at similar sites as are blue supergiants, and  
thus, the reason why their reddening is systematically higher is unclear. 
The direct determination of reddening from spectroscopy should be superior 
to the indirect way of using the difference of apparent distance moduli in $V$- 
and $I$-bands. We note that the latter must also depend on metallicity (an effect 
that had not been considered in the pertinent studies) as well as an
assumed value of reddening for the calibration  
sample of Cepheids in the LMC. If one assumes that the 
value of $E(B-V)=0.2$ mag as adopted for the Cepheids study is too high and 
uses the average value of 0.083 mag from the blue supergiants study
instead, then the Cepheid distance modulus would increase by 0.37 mag,
yielding a distance in excellent agreement with our FGLR and TRGB
values. This drastic improvement illustrates
two things: first, the HST observations of Cepheids seem to be compatible 
with a large distance modulus; second and most importantly, the accurate 
elimination of the reddening uncertainty is absolutely crucial for the 
determination of distances using Cepheids.

Very recently, \cite{Scowcroft09}
presented ground-based $B$-, $V$-, $I$-band photometry of 
Cepheids in the center of M\,33 and in the southern spiral arm, some 4 to 5 kpc 
away from the center, with 91 and 28 objects in the two fields, respectively. The 
use of the Wesenheit $W_{VI}$ magnitude, which was assumed to be reddening-free, 
yielded largely discrepant distance moduli in the center and in the spiral arm 
(24.37 and 24.54 mag, respectively) that the authors attributed to a 
metallicity difference between the regions. They derived a true distance 
modulus of $\mu$\,=\,24.53$\,\pm\,$0.11\,mag accordingly, very much in agreement with the 
HST photometry Cepheid papers discussed in the previous paragraph. The paper 
does not provide an estimate of $E(B-V)$, but given their distance agreement with the 
published HST Cepheid work, we assume that their reddening value would
be close to 0.2 mag. We also 
note that, at least from the basic principle, the use of the Wesenheit magnitude 
is very similar to the use of difference between apparent distance moduli 
obtained in $V$- and $I$-bands.  Therefore, we feel that the same
discussion on reddening and distance determination in the 
previous paragraphs applies here as well.  (This distance modulus is based on a
distance modulus of 18.40 mag for the LMC, while our FGLR calibration
adopted 18.50 mag.  Thus, it would increase by 0.1 mag if our distance
were assumed.)  

It is also important to take into account that the metallicity
difference between the two fields studied by \cite{Scowcroft09} seems
to be significantly smaller than the value of 0.566\,dex they adopted.  
This value was obtained based on the assumption of a two-component
slope as originally proposed by \cite{Vilchez88} and then
also adopted as one possibility by 
\citet[][]{Magrini07b}. However, as discussed in \S\ref{Metallicity}
and shown in our Fig.~\ref{fig:met_grad_magr}, both the metallicities
from the stars and from \ion{H}{2} regions do not support a very steep
gradient towards the center of M\,33.  Using our 
result from \S\ref{Metallicity} for the abundance gradient, the
metallicity difference between the two Scowcroft et al. fields is only
0.31 dex. With this number, the dependence of the Cepheid distance
moduli on logarithmic metallicity changes would become
$\gamma$\,=\,0.55 mag\,dex$^{-1}$, almost twice as high 
as the value of 0.29 found by \cite{Macri06} and that
derived by~\cite{Scowcroft09}.

There were also a number of other distance determinations methods in
the literature that involved stars of significantly older populations. The TRGB method, 
as applied in this work, is a typical example. Our distance modulus 
agrees very well with that of~\cite{Kim02}, who used HST WFPC photometry 
in 10 fields and obtained $\mu=24.81\,\pm\,$0.13 mag. Note also that 
\cite{Kim02} were the first to draw attention to the fact that the 
shorter distance modulus associated with Cepheid distances might be a result of 
overestimating of interstellar reddening.

In addition, three studies involving ground-based photometry were published 
later yielding quite a range in distance: \cite{McConnachie04}, 
$\mu=24.50$ $\pm$ 0.06 mag; \cite{Galleti04}, $\mu=24.64$ $\pm$ 0.15 mag; 
and \cite{Tiede04}, $\mu=24.69$ $\pm$ 0.07 mag. It seems that systematic 
effects such as $I$-band calibration and the algorithm used for TRGB edge 
detection are of importance when comparing these results. We note, however, 
that two of these studies agree with the results acquired with 
HST photometry within the corresponding error margins.

Other independent investigations using older stellar populations were carried 
out by~\cite{Sarajedini00} and~\cite{Ciardullo04} and 
confirmed a large distance modulus. The former studied Horizontal Branch stars 
and found $\mu=24.84$ $\pm$ 0.16 mag, whereas the latter used the Planetary 
Nebulae Luminosty Function to obtain $\mu=24.86^{+0.07}_{-0.11}$ mag.

\cite{Sarajedini06} used HST ACS photometry of RR Lyrae field stars 
because at minimum light, the intrinsic $V-I$ color of RR Lyrae stars 
is well defined and independent of metallicity and period. This allows for an
estimate of reddening from the construction of $V$ and $I$ light curves. Unfortunately, 
the reddening determination was somewhat uncertain and 18 stars in their sample 
turned out to have negative reddening. The authors found
$\mu=24.67$ $\pm$ 0.07 mag while excluding those objects, and
$\mu=24.76$ $\pm$ 0.08 mag with them included. We believe that these
18 stars are simply cases with very small reddening and represent the
uncertainty of the procedure; thus, their inclusion is
justified. This, then, exhibits agreement with the HST TRGB 
results as well as with that derived from the HB and PN stars.

In summary, we conclude that a large distance modulus as obtained with the 
FGLR and TRGB work presented here is well supported by the value of 
reddening we found and agrees well with other independent work based on low-mass stars of older 
populations.

 \section{Conclusions}
 \label{Conclusions}

Motivated by the need to achieve precision in extragalactic distance
determination and to resolve the large discrepancy in the distance
modulus to the nearby Triangulum Galaxy M33, we undertake an
independent investigation applying the flux-weighted gravity--luminosity
relationship to a sample of 22 blue supergiants at various galactocentric
distances in M33.  With medium resolution spectra from Keck, we carry
out a quantitative spectral analysis and measured stellar parameters
such as rotational velocities, stellar gravities, effective
temperatures, and metallicities.  Together with state-of-the-art
non-LTE model atmospheres spanning an expansive grid in 
($T_{\rm eff},~\log\,g$) parameter space and photometry available in
literature, we further derive interstellar reddening, stellar
luminosities, radii, and masses.  The combinations of these
parameters allow us to determine a distance modulus of
$24.93\,\pm\,0.11$ mag to M33. 

We discuss in detail the stellar properties of our supergiants in
relation to their predicted evolutionary paths.  On the
high-luminosity end, our spectroscopic results are nicely confirmed by
the evolutionary tracks on the H-R diagram.  On the low-luminosity end,
however, we observe one object (similar to the one found by K08)
that appears to follow a different 
path and offer close binary evolution as an explanation, which will
require further confirmation from detailed evolutionary calculations
in the future.  Furthermore, we note a small systematic discrepancy between
spectroscopic and evolutionary masses, however it is unlikely to be
caused by uncertainties in the distance modulus. Our discussion on
metallicities specifies the significant metallicity gradient as a
function of galactocentric distance that we find in M33, with
large scatter about the regression curve.  This is similar to the
results of several other metal abundance studies of \ion{H}{2} regions
in literature.

Our FGLR for M33 follows that for other galaxies very nicely, giving
confidence to both the reliability of its calibration and the distance
modulus thus determined.  Using HST photometry available in the
archive, we also present a TRGB distance modulus that nicely converges
with our FGLR result within the uncertainties.  These distances agree
well with other published large distance moduli and low reddening
values found in literature, based on work on detached eclipsing
binary, long period variables, horizontal branch stars, and planetary
nebulae luminosity function.  The discrepancy between our results
and that from Cepheid studies can be explained by differences in
interstellar reddening assumed in the different work, and we point out
that the Cepheid distance moduli will increase if the reddening value
and metallicity were properly determined and applied.
However, we note that a final conclusion will only be possible after a
more rigorous calibration of the FGLR using a large sample of LMC blue
supergiants.

\acknowledgements

The data presented herein were obtained
at the W.M. Keck Observatory, which is operated as a scientific
partnership among the California Institute of Technology, the
University of California and the National Aeronautics and Space
Administration. The Observatory was made possible by the generous
financial support of the W.M. Keck Foundation.  The authors wish to
recognize and acknowledge the very significant cultural role and
reverence that the summit of Mauna Kea has always had within the
indigenous Hawaiian community.  We are most fortunate to have the
opportunity to conduct observations from this mountain. 

\bibliography{bibliography}

\begin{thebibliography}{51}
\expandafter\ifx\csname natexlab\endcsname\relax\def\natexlab#1{#1}\fi

\bibitem[{{Allende Prieto} {et~al.}(2001){Allende Prieto}, {Lambert}, \&
  {Asplund}}]{Allende01}
{Allende Prieto}, C., {Lambert}, D.~L., \& {Asplund}, M. 2001, \apjl, 556, L63

\bibitem[{{Beaulieu} {et~al.}(2006){Beaulieu}, {Buchler}, {Marquette},
  {Hartman}, \& {Schwarzenberg-Czerny}}]{Beaulieu06}
{Beaulieu}, J.-P., {Buchler}, J.~R., {Marquette}, J.-B., {Hartman}, J.~D., \&
  {Schwarzenberg-Czerny}, A. 2006, \apjl, 653, L101

\bibitem[{{Bonanos} {et~al.}(2006){Bonanos}, {Stanek}, {Kudritzki}, {Macri},
  {Sasselov}, {Kaluzny}, {Stetson}, {Bersier}, {Bresolin}, {Matheson},
  {Mochejska}, {Przybilla}, {Szentgyorgyi}, {Tonry}, \& {Torres}}]{Bonanos06}
{Bonanos}, A.~Z., {et~al.} 2006, \apj, 652, 313

\bibitem[{{Bresolin} {et~al.}(2001){Bresolin}, {Kudritzki}, {Mendez}, \&
  {Przybilla}}]{Bresolin01}
{Bresolin}, F., {Kudritzki}, R.-P., {Mendez}, R.~H., \& {Przybilla}, N. 2001,
  \apjl, 548, L159

\bibitem[{{Brunthaler} {et~al.}(2005){Brunthaler}, {Reid}, {Falcke},
  {Greenhill}, \& {Henkel}}]{Brunthaler05}
{Brunthaler}, A., {Reid}, M.~J., {Falcke}, H., {Greenhill}, L.~J., \& {Henkel},
  C. 2005, Science, 307, 1440

\bibitem[{{Cardelli} {et~al.}(1989){Cardelli}, {Clayton}, \&
  {Mathis}}]{Cardelli89}
{Cardelli}, J.~A., {Clayton}, G.~C., \& {Mathis}, J.~S. 1989, \apj, 345, 245

\bibitem[{{Ciardullo} {et~al.}(2004){Ciardullo}, {Durrell}, {Laychak},
  {Herrmann}, {Moody}, {Jacoby}, \& {Feldmeier}}]{Ciardullo04}
{Ciardullo}, R., {Durrell}, P.~R., {Laychak}, M.~B., {Herrmann}, K.~A.,
  {Moody}, K., {Jacoby}, G.~H., \& {Feldmeier}, J.~J. 2004, \apj, 614, 167

\bibitem[{{de Vaucouleurs} {et~al.}(1995){de Vaucouleurs}, {de Vaucouleurs},
  {Corwin}, {Buta}, {Paturel}, \& {Fouque}}]{deVaucouleurs95}
{de Vaucouleurs}, G., {de Vaucouleurs}, A., {Corwin}, H.~G., {Buta}, R.~J.,
  {Paturel}, G., \& {Fouque}, P. 1995, VizieR Online Data Catalog, 7155, 0

\bibitem[{{Dolphin}(2000)}]{Dolphin00}
{Dolphin}, A.~E. 2000, \pasp, 112, 1383

\bibitem[{{Freedman} {et~al.}(1991){Freedman}, {Wilson}, \&
  {Madore}}]{Freedman91}
{Freedman}, W.~L., {Wilson}, C.~D., \& {Madore}, B.~F. 1991, \apj, 372, 455

\bibitem[{{Frieman} {et~al.}(2008){Frieman}, {Turner}, \&
  {Huterer}}]{Frieman08}
{Frieman}, J.~A., {Turner}, M.~S., \& {Huterer}, D. 2008, \araa, 46, 385

\bibitem[{{Galleti} {et~al.}(2004){Galleti}, {Bellazzini}, \&
  {Ferraro}}]{Galleti04}
{Galleti}, S., {Bellazzini}, M., \& {Ferraro}, F.~R. 2004, \aap, 423, 925

\bibitem[{{Gieren} {et~al.}(2005){Gieren}, {Pietrzy{\'n}ski}, {Soszy{\'n}ski},
  {Bresolin}, {Kudritzki}, {Minniti}, \& {Storm}}]{Gieren05}
{Gieren}, W., {Pietrzy{\'n}ski}, G., {Soszy{\'n}ski}, I., {Bresolin}, F.,
  {Kudritzki}, R.-P., {Minniti}, D., \& {Storm}, J. 2005, \apj, 628, 695

\bibitem[{{Gray}(1992)}]{Gray92}
{Gray}, D.~F. 1992, {The Observation and Analysis of Stellar Photospheres} (The
  Observation and Analysis of Stellar Photospheres, by David F.~Gray,
  pp.~470.~ISBN 0521408687.~Cambridge, UK: Cambridge University Press, June
  1992.)

\bibitem[{{Howarth}(2004)}]{Howarth04}
{Howarth}, I.~D. 2004, in IAU Symposium, Vol. 215, Stellar Rotation, ed.
  A.~{Maeder} \& P.~{Eenens}, 33--+

\bibitem[{{Humphreys} \& {Sandage}(1980)}]{Humphreys80}
{Humphreys}, R.~M., \& {Sandage}, A. 1980, \apjs, 44, 319

\bibitem[{{Ivanov} {et~al.}(1993){Ivanov}, {Freedman}, \& {Madore}}]{Ivanov93}
{Ivanov}, G.~R., {Freedman}, W.~L., \& {Madore}, B.~F. 1993, \apjs, 89, 85

\bibitem[{{Kim} {et~al.}(2002){Kim}, {Kim}, {Lee}, {Sarajedini}, \&
  {Geisler}}]{Kim02}
{Kim}, M., {Kim}, E., {Lee}, M.~G., {Sarajedini}, A., \& {Geisler}, D. 2002,
  \aj, 123, 244

\bibitem[{{Kudritzki} {et~al.}(2003){Kudritzki}, {Bresolin}, \&
  {Przybilla}}]{Kudritzki03}
{Kudritzki}, R.~P., {Bresolin}, F., \& {Przybilla}, N. 2003, \apjl, 582, L83

\bibitem[{Kudritzki {et~al.}(2008)Kudritzki, Urbaneja, Bresolin, Przybilla,
  Gieren, \& Pietrzynski}]{Kudritzki08}
Kudritzki, R.-P., Urbaneja, M.~A., Bresolin, F., Przybilla, N., Gieren, W., \&
  Pietrzynski, G. 2008, Astrophysical Journal, 681, 269

\bibitem[{{Lee} {et~al.}(2002){Lee}, {Kim}, {Sarajedini}, {Geisler}, \&
  {Gieren}}]{Lee02}
{Lee}, M.~G., {Kim}, M., {Sarajedini}, A., {Geisler}, D., \& {Gieren}, W. 2002,
  \apj, 565, 959

\bibitem[{{Lennon}(1997)}]{Lennon97}
{Lennon}, D.~J. 1997, \aap, 317, 871

\bibitem[{{Macri} {et~al.}(2006){Macri}, {Stanek}, {Bersier}, {Greenhill}, \&
  {Reid}}]{Macri06}
{Macri}, L.~M., {Stanek}, K.~Z., {Bersier}, D., {Greenhill}, L.~J., \& {Reid},
  M.~J. 2006, \apj, 652, 1133

\bibitem[{{Maeder} \& {Meynet}(2005)}]{Maeder05}
{Maeder}, A., \& {Meynet}, G. 2005, \aap, 440, 1041

\bibitem[{{Magrini} {et~al.}(2007{\natexlab{a}}){Magrini}, {Corbelli}, \&
  {Galli}}]{Magrini07a}
{Magrini}, L., {Corbelli}, E., \& {Galli}, D. 2007{\natexlab{a}}, \aap, 470,
  843

\bibitem[{{Magrini} {et~al.}(2007{\natexlab{b}}){Magrini}, {V{\'{\i}}lchez},
  {Mampaso}, {Corradi}, \& {Leisy}}]{Magrini07b}
{Magrini}, L., {V{\'{\i}}lchez}, J.~M., {Mampaso}, A., {Corradi}, R.~L.~M., \&
  {Leisy}, P. 2007{\natexlab{b}}, \aap, 470, 865

\bibitem[{{Makarov} {et~al.}(2006){Makarov}, {Makarova}, {Rizzi}, {Tully},
  {Dolphin}, {Sakai}, \& {Shaya}}]{Makarov06}
{Makarov}, D., {Makarova}, L., {Rizzi}, L., {Tully}, R.~B., {Dolphin}, A.~E.,
  {Sakai}, S., \& {Shaya}, E.~J. 2006, \aj, 132, 2729

\bibitem[{{Marinoni} {et~al.}(2001){Marinoni}, {Davis}, {Coil}, \&
  {Finkbeiner}}]{Marioni01}
{Marinoni}, C., {Davis}, M., {Coil}, A.~L., \& {Finkbeiner}, D. 2001, ArXiv
  Astrophysics e-prints

\bibitem[{{Massey} {et~al.}(1995){Massey}, {Armandroff}, {Pyke}, {Patel}, \&
  {Wilson}}]{Massey95}
{Massey}, P., {Armandroff}, T.~E., {Pyke}, R., {Patel}, K., \& {Wilson}, C.~D.
  1995, \aj, 110, 2715

\bibitem[{{Massey} {et~al.}(1996){Massey}, {Bianchi}, {Hutchings}, \&
  {Stecher}}]{Massey96}
{Massey}, P., {Bianchi}, L., {Hutchings}, J.~B., \& {Stecher}, T.~P. 1996,
  \apj, 469, 629

\bibitem[{{Massey} {et~al.}(2006){Massey}, {Olsen}, {Hodge}, {Strong},
  {Jacoby}, {Schlingman}, \& {Smith}}]{Massey06}
{Massey}, P., {Olsen}, K.~A.~G., {Hodge}, P.~W., {Strong}, S.~B., {Jacoby},
  G.~H., {Schlingman}, W., \& {Smith}, R.~C. 2006, \aj, 131, 2478

\bibitem[{{McConnachie} {et~al.}(2004){McConnachie}, {Irwin}, {Ferguson},
  {Ibata}, {Lewis}, \& {Tanvir}}]{McConnachie04}
{McConnachie}, A.~W., {Irwin}, M.~J., {Ferguson}, A.~M.~N., {Ibata}, R.~A.,
  {Lewis}, G.~F., \& {Tanvir}, N. 2004, \mnras, 350, 243

\bibitem[{{Meynet} \& {Maeder}(2003)}]{Meynet03}
{Meynet}, G., \& {Maeder}, A. 2003, \aap, 404, 975

\bibitem[{{Pierce} {et~al.}(2000){Pierce}, {Jurcevic}, \&
  {Crabtree}}]{Pierce00}
{Pierce}, M.~J., {Jurcevic}, J.~S., \& {Crabtree}, D. 2000, \mnras, 313, 271

\bibitem[{{Przybilla} {et~al.}(2006){Przybilla}, {Butler}, {Becker}, \&
  {Kudritzki}}]{Przybilla06}
{Przybilla}, N., {Butler}, K., {Becker}, S.~R., \& {Kudritzki}, R.~P. 2006,
  \aap, 445, 1099

\bibitem[{{Przybilla} {et~al.}(2008){Przybilla}, {Nieva}, \&
  {Butler}}]{Przybilla08}
{Przybilla}, N., {Nieva}, M.-F., \& {Butler}, K. 2008, \apjl, 688, L103

\bibitem[{{Puls} {et~al.}(2005){Puls}, {Urbaneja}, {Venero}, {Repolust},
  {Springmann}, {Jokuthy}, \& {Mokiem}}]{Puls05}
{Puls}, J., {Urbaneja}, M.~A., {Venero}, R., {Repolust}, T., {Springmann}, U.,
  {Jokuthy}, A., \& {Mokiem}, M.~R. 2005, \aap, 435, 669

\bibitem[{{Rizzi} {et~al.}(2007){Rizzi}, {Tully}, {Makarov}, {Makarova},
  {Dolphin}, {Sakai}, \& {Shaya}}]{Rizzi07}
{Rizzi}, L., {Tully}, R.~B., {Makarov}, D., {Makarova}, L., {Dolphin}, A.~E.,
  {Sakai}, S., \& {Shaya}, E.~J. 2007, \apj, 661, 815

\bibitem[{{Rosolowsky} \& {Simon}(2008)}]{Rosolowsky08}
{Rosolowsky}, E., \& {Simon}, J.~D. 2008, \apj, 675, 1213

\bibitem[{{Rubin} {et~al.}(2008){Rubin}, {Simpson}, {Colgan}, {Dufour},
  {Brunner}, {McNabb}, {Pauldrach}, {Erickson}, {Haas}, \& {Citron}}]{Rubin08}
{Rubin}, R.~H., {et~al.} 2008, \mnras, 387, 45

\bibitem[{{Sarajedini} {et~al.}(2006){Sarajedini}, {Barker}, {Geisler},
  {Harding}, \& {Schommer}}]{Sarajedini06}
{Sarajedini}, A., {Barker}, M.~K., {Geisler}, D., {Harding}, P., \& {Schommer},
  R. 2006, \aj, 132, 1361

\bibitem[{{Sarajedini} {et~al.}(2000){Sarajedini}, {Geisler}, {Schommer}, \&
  {Harding}}]{Sarajedini00}
{Sarajedini}, A., {Geisler}, D., {Schommer}, R., \& {Harding}, P. 2000, \aj,
  120, 2437

\bibitem[{{Schiller} \& {Przybilla}(2008)}]{Schiller08}
{Schiller}, F., \& {Przybilla}, N. 2008, \aap, 479, 849

\bibitem[{{Schlegel} {et~al.}(1998){Schlegel}, {Finkbeiner}, \&
  {Davis}}]{Schlegel98}
{Schlegel}, D.~J., {Finkbeiner}, D.~P., \& {Davis}, M. 1998, \apj, 500, 525

\bibitem[{{Scowcroft} {et~al.}(2009){Scowcroft}, {Bersier}, {Mould}, \&
  {Wood}}]{Scowcroft09}
{Scowcroft}, V., {Bersier}, D., {Mould}, J.~R., \& {Wood}, P.~R. 2009, \mnras,
  635

\bibitem[{{Sheinis} {et~al.}(2002){Sheinis}, {Bolte}, {Epps}, {Kibrick},
  {Miller}, {Radovan}, {Bigelow}, \& {Sutin}}]{Sheinis02}
{Sheinis}, A.~I., {Bolte}, M., {Epps}, H.~W., {Kibrick}, R.~I., {Miller},
  J.~S., {Radovan}, M.~V., {Bigelow}, B.~C., \& {Sutin}, B.~M. 2002, \pasp,
  114, 851

\bibitem[{{Tiede} {et~al.}(2004){Tiede}, {Sarajedini}, \& {Barker}}]{Tiede04}
{Tiede}, G.~P., {Sarajedini}, A., \& {Barker}, M.~K. 2004, \aj, 128, 224

\bibitem[{{Urbaneja} {et~al.}(2005{\natexlab{a}}){Urbaneja}, {Herrero},
  {Bresolin}, {Kudritzki}, {Gieren}, {Puls}, {Przybilla}, {Najarro}, \&
  {Pietrzy{\'n}ski}}]{Urbaneja05b}
{Urbaneja}, M.~A., {et~al.} 2005{\natexlab{a}}, \apj, 622, 862

\bibitem[{{Urbaneja} {et~al.}(2005{\natexlab{b}}){Urbaneja}, {Herrero},
  {Kudritzki}, {Najarro}, {Smartt}, {Puls}, {Lennon}, \& {Corral}}]{Urbaneja05}
{Urbaneja}, M.~A., {Herrero}, A., {Kudritzki}, R.-P., {Najarro}, F., {Smartt},
  S.~J., {Puls}, J., {Lennon}, D.~J., \& {Corral}, L.~J. 2005{\natexlab{b}},
  \apj, 635, 311

\bibitem[{{Urbaneja} {et~al.}(2008){Urbaneja}, {Kudritzki}, {Bresolin},
  {Przybilla}, {Gieren}, \& {Pietrzy{\'n}ski}}]{Urbaneja08}
{Urbaneja}, M.~A., {Kudritzki}, R.-P., {Bresolin}, F., {Przybilla}, N.,
  {Gieren}, W., \& {Pietrzy{\'n}ski}, G. 2008, \apj, 684, 118

\bibitem[{{Vilchez} {et~al.}(1988){Vilchez}, {Pagel}, {Diaz}, {Terlevich}, \&
  {Edmunds}}]{Vilchez88}
{Vilchez}, J.~M., {Pagel}, B.~E.~J., {Diaz}, A.~I., {Terlevich}, E., \&
  {Edmunds}, M.~G. 1988, \mnras, 235, 633

\end{thebibliography}
\bibliographystyle{apj}

\clearpage

  \begin{deluxetable}{ c c l l }
    \centering
    \tablewidth{0pt}
    \tablecolumns{3}
    \tablecaption{Target Identification}
    \tablehead{   
      \colhead{No.}           &
      \colhead{Massey et al. (2006) ID} &
      \colhead{Alt. ID}      & 
      \colhead{Reference}
	}
    \startdata
 1 & J013351.20+303224.5 &               & \\  
 2 & J013337.09+303521.6 &               & \\  
 3 & J013340.47+303503.3 &               & \\  
 4 & J013340.84+303822.5 &               & \\  
 5 & J013341.36+303629.6 &               & \\  
 6 & J013344.27+304247.2 &               & \\  
 7 & J013344.43+303843.9 &               & \\  
 8 & J013344.81+303217.8 &               & \\  
 9 & J013340.55+303158.7 &               & \\  
10 & J013343.26+303153.8 &              & \\  
11 & J013340.30+302144.1 & 0755       &  1 \\  
12 & J013344.66+303631.6 & B215a     &  2 \\  
13 & J013229.61+303513.3 &                & \\  
14 & J013315.62+302949.3 &                & \\  
15 & J013351.56+304005.2 &                & \\  
16 & J013300.23+302323.7 & 117\,A     & 2 \\  
17 & J013339.08+302010.7 &                & \\  
18 & J013300.86+303504.9 & B38, OB\,21-108, UIT\.030       & 2, 3, 4    \\
19 & J013342.06+302142.3 & B287, OB\,112-41 & 2, 3 \\
20 & J013335.76+310046.9 & B157, UIT\,136     & 2, 4 \\
21 & J013333.72+304719.9 & UIT\,122                 & 4 \\
22 & J013327.35+310056.4 & UIT\,103                 & 4 \\
    \enddata						 
    \label{tbl:id}	
   \tablecomments{ References: (1) \citet{Ivanov93}, (2) \citet{Humphreys80}, 
   (3) \citet{Massey95}, (4) \citet{Massey96} }   				 
  \end{deluxetable}

  \begin{deluxetable}{ccccccccccc}
    \centering
    \setlength{\tabcolsep}{0.04in} 
    \tablewidth{0pt}
    \tablecolumns{10}
    \tablecaption{Spectra, Spectral Types, and Spectroscopic Parameters}
    \tablehead{   
      \colhead{No.} &
      \colhead{Sp.} &
      \colhead{S\,\tablenotemark{1}} &
      \colhead{$v\,\sin\,i$} &      
      \colhead{$\zeta$\,\tablenotemark{2}} &
      \colhead{$R/R_{25}$\,\tablenotemark{3} } &
      \colhead{T\,\tablenotemark{4}} &
      \colhead{$T_{\rm eff}$} &
      \colhead{$\log~g$} &
      \colhead{$[Z]$} &
      \colhead{$\log~g_F$} \\
      \colhead{ } &
      \colhead{Type} &
      \colhead{ } &
      \colhead{(\kms)} &
      \colhead{(\kms)} &
      \colhead{ } &
      \colhead{ } &
      \colhead{(K)} &
      \colhead{ (dex) } &
      \colhead{ (dex) } &
      \colhead{ (dex) }
	}
    \startdata
     1 & B9 & D & 40 &       & 0.23 & b & 11000$\pm$500 & 1.55$\pm$0.14 & -0.07$\pm$0.15 & 1.38$\pm$0.06\\
     2 & A2 & D & 27 &       & 0.15 & a & 8500$\pm$225 & 1.15$\pm$0.15 & 0.01$\pm$0.12 & 1.43$\pm$0.10\\
     3 & B9 & D & 53 &       & 0.14 & b & 11500$\pm$500 & 1.70$\pm$0.13 & 0.08$\pm$0.20 & 1.46$\pm$0.06\\
     4 & A2 & D & 25 &       & 0.09 & a & 8500$\pm$225 & 1.50$\pm$0.20 & 0.23$\pm$0.15 & 1.78$\pm$0.06\\
     5 & A2 & D & 22 &       & 0.11 & a & 8750$\pm$250 & 1.30$\pm$0.18 & 0.10$\pm$0.15 & 1.53$\pm$0.13\\
     6 & B3 & D & 45 &       & 0.14 & c & 16000$\pm$1000 & 2.15$\pm$0.20 & 0.00$\pm$0.10 & 1.33$\pm$0.09\\
     7 & B9 & D & 30 &       & 0.06 & b & 11000$\pm$500 & 1.65$\pm$0.15 & 0.12$\pm$0.23 & 1.48$\pm$0.08\\ 
     8 & B8 & D & 56 &       & 0.22 & b & 12500$\pm$500 & 1.90$\pm$0.12 & -0.01$\pm$0.12 & 1.51$\pm$0.05\\ 
     9 & B3 & D & 58 &       & 0.23 & c & 16000$\pm$1000 & 2.15$\pm$0.10 & 0.00$\pm$0.10 & 1.33$\pm$0.09\\ 
    10 & A0 & D & 42 &      & 0.23 & a,b & 10000$\pm$500 & 1.35$\pm$0.15 & -0.02$\pm$0.20 & 1.35$\pm$0.06\\ 
    11 & A0 & E & 35 & 30 & 0.54 & a,b & 9600$\pm$340 & 1.38$\pm$0.11 & -0.50$\pm$0.18 & 1.45$\pm$0.05\\  
    12 & A0 & E & 50 & 30 & 0.09 & a,b & 9700$\pm$270 & 1.32$\pm$0.12 & -0.05$\pm$0.21 & 1.37$\pm$0.07\\
    13 & A0 & E & 42 & 20 & 0.76 & a,b & 9550$\pm$320 & 1.41$\pm$0.12 & -0.48$\pm$0.19 & 1.49$\pm$0.06\\
    14 & B9 & E & 35 & 25 & 0.37 & a,b & 10000$\pm$410 & 1.32$\pm$0.16 & -0.37$\pm$0.21 & 1.32$\pm$0.09\\
    15 & B9 & E & 45 & 30 & 0.01 & a,b & 11500$\pm$500 & 1.63$\pm$0.16 & 0.03$\pm$0.22 & 1.39$\pm$0.08\\
    16 & A0 & E & 45 & 30 & 0.57 & a,b & 9600$\pm$328 & 1.10$\pm$0.12 & -0.52$\pm$0.19 & 1.17$\pm$0.06\\
    17 & B3 & E & 50 & 40 & 0.59 & c & 17000$\pm$1000 & 2.00$\pm$0.15 & -0.17$\pm$0.20 & 1.08$\pm$0.05\\
    18\tablenotemark{5} & B1 & I      & 60 &    & 0.46 & c & 22000$\pm$500 & 2.60$\pm$0.10 & -0.22$\pm$0.15 & 1.23$\pm$0.14\\      
    19\tablenotemark{5} & B0 & E    & 65 &    & 0.54 & c & 26000$\pm$500 & 2.79$\pm$0.10 & 0.00$\pm$0.15 & 1.13$\pm$0.13\\       
    20\tablenotemark{5} & B0.5 & E & 70 &    & 0.74 & c & 24000$\pm$500 & 2.60$\pm$0.10 & -0.30$\pm$0.15 & 1.08$\pm$0.13\\   
    21\tablenotemark{5} & B0.5 & E & 70 &    & 0.35 & c & 24000$\pm$500 & 2.77$\pm$0.10 & -0.22$\pm$0.15 & 1.25$\pm$0.13\\     
    22\tablenotemark{5} & B0.7 & E & 50 &    & 0.79 & c & 23700$\pm$500 & 2.65$\pm$0.10 & -0.52$\pm$0.15 & 1.15$\pm$0.13
    \enddata
    \label{tbl:summary}
    \tablenotetext{1}{Spectrograph: D -- DEIMOS, E -- ESI, I -- ISIS}
    \tablenotetext{2}{Macroturbulence velocity}
    \tablenotetext{3}{$R_{25}\,=\,35.40^\prime$, the radius of M33
	in the plane of the galaxy; this has been corrected for a position
	angle of $22^o$ and an inclination angle of $54^o$ }
    \tablenotetext{4}{$T_{\rm eff}$-method. a: Mg~{\sc i/ii}, b: \ion{He}{1}, c: Si~{\sc ii/iii/iv} }
    \tablenotetext{5}{Parameters adopted from \cite{Urbaneja05}, with the galactocentric distances corrected for the normalization radius and the distance to the galaxy used in this paper}
  \end{deluxetable}

  \begin{deluxetable}{ccccccc}
    \centering
    \setlength{\tabcolsep}{0.04in} 
    \tablewidth{0pt}
    \tablecolumns{7}
    \tablecaption{Photometric Properties and Bolometric Corrections}
    \tablehead{   
      \colhead{No.} &
      \colhead{$m_V$\,\tablenotemark{1}} &
      \colhead{$B-V$\,\tablenotemark{1}} &
      \colhead{$(B-V)_0$} &
      \colhead{$E(B-V)$} &
      \colhead{$M_V$\,\tablenotemark{2}} &
      \colhead{$BC$} \\
      \colhead{ } &
      \colhead{(mag)} &
      \colhead{(mag)} &
      \colhead{(mag)} &
      \colhead{(mag)} &
      \colhead{(mag)} &
      \colhead{(mag)} 
	}
    \startdata
     1 & 17.194$\pm$0.004 &  0.034$\pm$0.004 & -0.046$\pm$0.021 & 0.080$\pm$0.021 & -7.984$\pm0.065$ & -0.481$\pm$0.082 \\
     2 & 17.937$\pm$0.004 &  0.062$\pm$0.004 &  0.042$\pm$0.032 & 0.020$\pm$0.032 & -7.055$\pm0.099$ &  0.005$\pm$0.030 \\
     3 & 17.997$\pm$0.004 &  0.025$\pm$0.004 & -0.067$\pm$0.020 & 0.092$\pm$0.020 & -7.218$\pm0.062$ & -0.555$\pm$0.085 \\
     4 & 18.235$\pm$0.003 &  0.172$\pm$0.004 &  0.012$\pm$0.025 & 0.160$\pm$0.025 & -7.191$\pm0.078$ &  0.025$\pm$0.045 \\
     5 & 17.913$\pm$0.004 &  0.106$\pm$0.004 &  0.019$\pm$0.028 & 0.087$\pm$0.028 & -7.284$\pm0.087$ & -0.032$\pm$0.032 \\
     6 & 18.209$\pm$0.004 &  0.011$\pm$0.004 & -0.131$\pm$0.022 & 0.142$\pm$0.022 & -7.161$\pm0.068$ & -1.343$\pm$0.139 \\
     7 & 17.941$\pm$0.003 &  0.054$\pm$0.003 & -0.055$\pm$0.020 & 0.109$\pm$0.020 & -7.327$\pm0.062$ & -0.458$\pm$0.085 \\
     8 & 18.067$\pm$0.004 &  0.025$\pm$0.004 & -0.093$\pm$0.013 & 0.118$\pm$0.014 & -7.229$\pm0.044$ & -0.740$\pm$0.087 \\
     9 & 18.047$\pm$0.004 & -0.039$\pm$0.004 & -0.131$\pm$0.010 & 0.090$\pm$0.011 & -7.162$\pm0.034$ & -1.343$\pm$0.070 \\
    10 & 17.221$\pm$0.004 &  0.046$\pm$0.004 & -0.011$\pm$0.023 & 0.057$\pm$0.023 & -7.886$\pm0.071$ & -0.301$\pm$0.081 \\
    11 & 17.856$\pm$0.004 &  0.044$\pm$0.004 & -0.016$\pm$0.020 & 0.060$\pm$0.020 & -7.260$\pm0.032$ & -0.240$\pm$0.050 \\ 
    12 & 16.904$\pm$0.004 &  0.031$\pm$0.004 & -0.024$\pm$0.020 & 0.055$\pm$0.020 & -8.197$\pm0.032$ & -0.240$\pm$0.050 \\
    13 & 17.623$\pm$0.004 &  0.108$\pm$0.004 & -0.020$\pm$0.009 & 0.130$\pm$0.010 & -7.710$\pm0.031$ & -0.230$\pm$0.040 \\ 
    14 & 17.120$\pm$0.004 &  0.061$\pm$0.004 & -0.019$\pm$0.030 & 0.080$\pm$0.030 & -8.058$\pm0.093$ & -0.330$\pm$0.050 \\
    15 & 17.102$\pm$0.003 &  0.039$\pm$0.003 & -0.081$\pm$0.030 & 0.120$\pm$0.030 & -8.200$\pm0.093$ & -0.570$\pm$0.060 \\
    16 & 16.440$\pm$0.004 &  0.142$\pm$0.004 & -0.012$\pm$0.020 & 0.130$\pm$0.030 & -8.893$\pm0.093$ & -0.300$\pm$0.020 \\
    17 & 17.209$\pm$0.005 & -0.075$\pm$0.005 & -0.150$\pm$0.019 & 0.075$\pm$0.020 & -7.954$\pm0.032$ & -1.490$\pm$0.140 \\
    18 & 17.322$\pm$0.004 & -0.118$\pm$0.004 & -0.170$\pm$0.010 & 0.052$\pm$0.011 & -7.769$\pm0.034$ & -2.110$\pm$0.083 \\
    19 & 17.989$\pm$0.004 & -0.179$\pm$0.004 & -0.190$\pm$0.010 & 0.012$\pm$0.011 & -6.975$\pm0.034$ & -2.470$\pm$0.084 \\
    20 & 17.898$\pm$0.005 & -0.144$\pm$0.005 & -0.170$\pm$0.010 & 0.026$\pm$0.011 & -7.113$\pm0.034$ & -2.340$\pm$0.084 \\
    21 & 17.754$\pm$0.004 & -0.063$\pm$0.004 & -0.180$\pm$0.005 & 0.117$\pm$0.006 & -7.539$\pm0.019$ & -2.340$\pm$0.084 \\
    22 & 18.035$\pm$0.005 & -0.154$\pm$0.005 & -0.180$\pm$0.005 & 0.026$\pm$0.007 & -6.976$\pm0.022$ & -2.320$\pm$0.084 
    \enddata						 
    \label{tbl:photo}					 
    \tablenotetext{1}{Adopted from \cite{Massey06}}
    \tablenotetext{2}{Computed with distance modulus of 24.93$\,\pm\,$0.11\,mag (this work) }
  \end{deluxetable}

  \begin{deluxetable}{rccrrr}
    \centering
    \tablewidth{0pt}
    \tablecolumns{6}
    \tablecaption{Bolometric Magnitudes, Luminosities, and Stellar Masses}
    \tablehead{   
      \colhead{No.} &
      \colhead{$M_{bol}$\,\tablenotemark{1}} &
      \colhead{$\log\,\left(L/L_{\odot}\right)$} &
      \colhead{$R$} &
      \colhead{$M_\mathrm{spec}$} &
      \colhead{$M_\mathrm{evol}$}  \\
     \colhead{ } &
      \colhead{(mag)} &
      \colhead{ (dex) } &
      \colhead{($R_{\odot}$)} &      
      \colhead{($M_{\odot}$)} &
      \colhead{($M_{\odot}$)}  
      }
    \startdata
     1 & -8.51$\pm$0.11 & 5.30$\pm$0.04 & 121$\pm$13 & 19.0$\pm$8.8 & 20.2$\pm$0.7 \\
     2 & -7.05$\pm$0.10 & 4.73$\pm$0.04 & 106$\pm$8 &  5.8$\pm$3.2 & 13.1$\pm$0.4 \\
     3 & -7.77$\pm$0.11 & 5.01$\pm$0.04 &  81$\pm$8 &  11.9$\pm$5.1 & 16.2$\pm$0.5 \\
     4 & -7.17$\pm$0.09 & 4.77$\pm$0.04 & 112$\pm$8 &  14.3$\pm$8.8 & 13.6$\pm$0.3 \\
     5 & -7.32$\pm$0.09 & 4.83$\pm$0.04 & 113$\pm$8 &   9.3$\pm$5.1 & 14.2$\pm$0.4 \\
     6 & -8.50$\pm$0.16 & 5.30$\pm$0.06 &  58$\pm$9 &  17.5$\pm$12.5 & 20.9$\pm$1.0 \\
     7 & -7.79$\pm$0.11 & 5.02$\pm$0.04 &  89$\pm$10 & 12.8$\pm$6.3 & 16.3$\pm$0.5 \\
     8 & -7.97$\pm$0.11 & 5.09$\pm$0.04 &  75$\pm$7 & 16.1$\pm$6.3 & 17.2$\pm$0.5 \\
     9 & -8.54$\pm$0.08 & 5.32$\pm$0.03 &  58$\pm$4 & 17.5$\pm$5.4 & 20.9$\pm$0.5 \\
    10 & -8.19$\pm$0.11 & 5.18$\pm$0.04 & 129$\pm$15 & 13.6$\pm$6.9 & 18.4$\pm$0.6 \\
    11 & -7.51$\pm$0.08 & 4.90$\pm$0.03 & 103$\pm$9 & 9.2$\pm$3.2 & 15.0$\pm$0.3 \\ 
    12 & -8.44$\pm$0.08 & 5.28$\pm$0.03 & 154$\pm$11 & 18.0$\pm$6.5 & 20.0$\pm$0.5 \\
    13 & -7.94$\pm$0.05 & 5.08$\pm$0.02 & 127$\pm$9 & 15.0$\pm$5.4 & 17.1$\pm$0.3 \\ 
    14 & -8.39$\pm$0.11 & 5.26$\pm$0.04 & 142$\pm$14 & 15.2$\pm$7.8 & 19.7$\pm$0.7 \\
    15 & -8.77$\pm$0.11 & 5.41$\pm$0.04 & 128$\pm$13 & 25.4$\pm$13.2 & 22.4$\pm$0.8 \\
    16 & -9.19$\pm$0.07 & 5.58$\pm$0.03 & 223$\pm$17 & 22.8$\pm$8.4 & 26.0$\pm$0.6 \\
    17 & -9.44$\pm$0.15 & 5.68$\pm$0.06 &  80$\pm$12 & 23.1$\pm$12.7 & 29.3$\pm$1.7 \\
    18 & -9.88$\pm$0.09 & 5.85$\pm$0.04 &  58$\pm$4 & 48.9$\pm$14.6 & 35.2$\pm$1.4 \\
    19 & -9.45$\pm$0.09 & 5.68$\pm$0.04 &  34$\pm$2 & 26.1$\pm$7.6 & 29.3$\pm$1.0 \\
    20 & -9.45$\pm$0.09 & 5.68$\pm$0.04 &  40$\pm$2 & 23.3$\pm$6.9 & 29.4$\pm$1.0 \\
    21 & -9.88$\pm$0.09 & 5.85$\pm$0.04 &  49$\pm$3 & 51.1$\pm$15.0 & 35.2$\pm$1.3 \\
    22 & -9.30$\pm$0.09 & 5.62$\pm$0.04 &  38$\pm$2 & 23.8$\pm$7.0 & 27.7$\pm$0.9 
    \enddata						 
    \label{tbl:masses}					 
    \tablenotetext{1}{Computed with distance modulus of 24.93$\,\pm\,$0.11\,mag (this work) }
  \end{deluxetable}

\end{document}